\documentclass[11pt]{article}
\usepackage{epsfig,psfig,float,amssymb}
\def\gsim{\;\raise0.3ex\hbox{$>$\kern-0.75em\raise-1.1ex\hbox{$\sim$}}\;}
\def\lsim{\;\raise0.3ex\hbox{$<$\kern-0.75em\raise-1.1ex\hbox{$\sim$}}\;}

\begin{document}

\begin{center}
{\large{\bf $\displaystyle{\frac{N}{D}}$ Description of Two Meson Amplitudes 
and Chiral Symmetry}}
\end{center}
\vspace{1cm}

\begin{center}
{\large{J.A. Oller and E. Oset}}
\end{center}

\begin{center}   
{\small{\it Departamento de F\'{\i}sica Te\'orica and IFIC\\
Centro Mixto Universidad de Valencia-CSIC\\
46100 Burjassot (Valencia), Spain}}
\end{center}

\vspace{3cm}
\begin{abstract}
{\small{The most general structure of an elastic partial wave amplitude
when the unphysical cuts are neglected is deduced in terms of the N/D method. 
This result is then matched to lowest order, ${\mathcal{O}}(p^2)$, Chiral 
Perturbation Theory($\chi$PT) and to the exchange (consistent with chiral symmetry) 
 of resonances in the s-channel. The extension 
of the method to coupled channels is also given. Making use of the former
formalism, the $\pi\pi$ and 
$K\pi$(I=1/2) P-wave scattering amplitudes are described without free 
parameters when taking into 
account relations coming from the 1/$N_c$ expansion 
and unitarity. 
Next, the scalar sector is studied and good agreement with experiment up to 
$\sqrt{s}=1.4$ GeV is found. It is observed that the $a_0(980)$, $\sigma$ and 
$\kappa(900)$ resonances are meson-meson states originating from the 
unitarization of the ${\mathcal{O}}(p^2)$ $\chi$PT 
amplitudes. On the other hand, the $f_0(980)$ is a combination of a strong
S-wave meson-meson unitarity effect and of a preexisting singlet resonance with a mass 
around 1 GeV. We have also studied the size of the contributions of the 
unphysical cuts to the $\pi\pi$(I=0) and $K\pi$(I=1/2) elastic S-wave 
amplitudes from $\chi$PT and the exchange of resonances in crossed
channels up to $\sqrt{s}\approx 800$ MeV. The loops are calculated as in 
$\chi$PT at next to leading order.We find a small correction from the 
unphysical cuts to our calculated partial waves.}} 
\end{abstract}

\newpage

\section{Introduction}

       The understanding of the scalar meson-meson strong interaction is 
still so controversial that there is not even a consensus about how many low lying 
states (with masses $\leq$1 GeV) are there. The main difficulties which 
appear in this sector are: First, the possible presence of large width resonances, 
as the $f_0(400-1200)\equiv \sigma$ in $\pi\pi$ scattering or the 
$K^*_0(900)\equiv \kappa$ in the 
I=1/2 $K\pi$ amplitude, which cannot be easily distinguished from background
 contributions. 
Second, the existence of some resonances which appear just in the opening 
of an important channel with which they couple strongly, as for example 
the $f_0(980)$ or the $a_0(980)$ with the $K\bar{K}$ threshold around 1 GeV.
 All these aspects 
make that, for instance, it is not clear how many states are present, which is 
their nature and why simple parameterization of the scalar physical 
amplitudes in terms of standard Breit-Wigner resonances are not adequate, 
as stressed in several works \cite{Au,Tornqvist,Achasov2}. 

The conflictive situation for the scalar sector contrasts with the much 
better understood vector channels. In this latter case, one can 
achieve a sound understanding of the physics involved just from first 
principles \cite{FP}, namely, chiral symmetry, unitarity and relations 
coming from the QCD limit of infinity numbers of colors (Large $N_c$ QCD). This
is accomplished thanks to the leading role of the $\rho(770)$ and $K^*(890)$ 
resonances, in accordance with vector meson dominance, very well established from
particle and nuclear physics phenomenology. The issue is whether such a basic 
understanding for the scalar sector is possible, and, at the same time, 
is able to reproduce the associated phenomenology. Connected with the 
former, it should be also interesting to see if some kind of scalar meson 
dominance remains, in analogy with the above mentioned vector meson dominance.

There have been many studies of the scalar sector but none in the basic 
lines we outlined before. These studies have led to a 
variety of models dealing with the scalar meson-meson interaction and its 
associated low energy spectroscopy. The low energy scalar states
  have been ascribed \cite{Montanet} to conventional $q\bar{q}$ 
  mesons \cite{Morgan,Tornqvist}, $q^2\bar{q}^2$ states \cite{Jaffe,Achasov},
  $K\bar{K}$ molecules \cite{Weinstein,Jansen}, glueballs \cite{5deMP93} 
  and/or hybrids \cite{6deMP93}.
 
One can think in two possible ways in order to avoid the former explicit model 
dependence for the scalar sector, apart, of course, of solving QCD in four 
dimensions for low energies which, nowadays, is not affordable.

One way to proceed is to make use only of general principles that the physical 
amplitudes must fulfill, as unitarity and analyticity. There are a series 
of works by Pennington, Morgan et al. \cite{Au,Penotros} 
which fit nicely the experimental data available for the scalar-isoscalar 
sector and 
try to obtain also some understanding of the associated spectroscopy. 
Another work in this line is presented in \cite{Zug}. However, these approaches have also 
problems as, for example, the  specific way in which the amplitudes are 
parametrized and the lack of enough precision in the experimental data in 
order to discard other possible solutions.

Another alternative is the use of effective field theories which embody and 
exploit the symmetries of the underlying dynamics, in this case QCD. In this 
sense, Chiral Perturbation Theory ($\chi PT$) \cite{Weinberg,Retaila}, is the 
effective field theory of QCD with the lightest three quark flavors. This 
approach has been extensively used in the last years for the meson sector 
and allows to calculate any physical amplitude in a systematic power momentum 
expansion. 

This latter point of view will be the one adopted here. First, we will derive, 
making use of the N/D method \cite{Chew1}, the most general structure for an 
arbitrary partial wave amplitude when the unphysical cuts are neglected. In 
this way, our method can be seen as the zero order approach to a partial wave 
when treating the unphysical cuts in a perturbative sense. We think 
that this will be the case, at least, in those partial wave amplitudes which 
are dominated by unitarity and the presence of resonances in the s-channel with 
the same quantum numbers of the partial wave amplitude. For example, the 
case of the $\rho$ and $K^*$ resonances in the P wave $\pi\pi$ and $K\pi$ 
scattering respectively or the scalar channels with isospin 0, 1/2 
or 1, where several resonances appear. In fact, for the $\rho$($K^*$)
meson-meson channels, at least up to $\sqrt{s}=1.2$ GeV, one can describe
accurately the associated $\pi\pi$($K\pi$) phase shifts just in terms of simple 
Breit-Wigner parameterizations with the coupling of the $\rho$($K^*$) with 
$\pi\pi$($K\pi$) given by the KSFR relation \cite{KSFR} and their masses taken 
directly by experiment. In this way one has a free parameter description for
these processes without the unphysical cuts, since it only has the physical or 
right hand one as required by unitarity. Thus, one can deduce that for these
processes the contribution from the unphysical cuts is certainly much smaller
than the one coming from the exchange of these resonances in the s-channel and
from unitarity. Otherwise, a free parameter reproduction of these channels with
only the right hand cut could not be possible. This type of description for the 
$\rho$ and $K^*$ meson-meson channels is given below in {\bf{section 3}} and it 
was also given in ref. \cite{FP} for the case of the $\rho$. For the scalar
channels with I=0,1 and 1/2 there have been a number of previous studies 
\cite{Tornqvist,Achasov2,Achasov,48,JRlargo} which neglect the contribution from the
unphysical cuts establishing clearly the great importance of unitarity for the
scalar sector. In particular, in the work of ref. \cite{48} the I=0,1 S-wave
channels were described in terms of just one free parameter up to
$\sqrt{s}\approx 1.2$ GeV, indicating that the contribution from the unphysical
cuts should be small enough to be reabsorbed in this free parameter. The
connection between the work of ref. \cite{48} and the present one is discussed
in {\bf{section 4}} before the subsection dedicated to the resonance content of
our S-wave amplitudes. It is also interesting to indicate that in
\cite{RD1,FO1} the S-wave scattering was also studied including the unphysical
cuts up to ${\mathcal{O}}(p^4)$ in $\chi PT$ and the results obtained were very
similar to the ones of the former works, refs. \cite{48,JRlargo}, without any 
unphysical cuts at all. Apart from these considerations, we approach in the 
last section of the present work the influence of the unphysical cuts from 
$\chi$PT and the exchange of resonances in crossed channels taking into account 
the results of ref. \cite{UMR}. In this reference, the $\pi\pi$ 
and $K\pi$ elastic amplitudes are calculated up to one loop including explicit 
resonance fields \cite{EP}. In this way the range of applicability of 
$\chi$PT is extended up to
$\sqrt{s}\approx 700-800$ MeV. The loops are calculated as in $\chi$PT at 
$\mathcal{O}(p^4)$. We conclude that the contributions of the unphysical cuts 
are small and soft enough to be reabsorbed in our free parameters in a 
convergent way when treating the left hand cut in a perturbative way. It is 
important to indicate that such a small value for the influence of the 
unphysical cuts, as can be see in Table 3, is due to a
cancellation between the contributions coming from the loops and the exchange 
of resonances in crossed channels. 

 After neglecting the unphysical cuts we then match the 
general structure we obtain from the N/D method with the lowest order 
$\chi PT$ Lagrangian, ${\mathcal{O}}(p^2)$ \cite{Retaila}, and its extension 
to include heavier 
meson states with spin $\leq$ 1 \cite{EP}, beyond the lightest pseudoscalars 
($\pi,K,\eta$). Making use of this final formalism, we will study the scalar 
sector, being able to reproduce 
the experimental data up to about $\sqrt{s} \leq 1.4$ GeV, with $s$ the 
Mandelstam variable corresponding to the square of the total momentum of 
the pair of mesons. In order to say something for higher energies, 
more channels, apart from the ones taken here, should be added. This is 
not considered in this work, although it can be done in a straightforward way, 
albeit cumbersome, in terms of the present formalism. We also study the 
vector $\pi\pi$ and $K\pi$(I=1/2) scattering and compare it with the scalar 
sector to illustrate some important differences between both cases.

The main conclusion of the work is that one can obtain a rather accurate 
description of the scalar sector compared to experiment, in a way consistent 
with $\chi PT$ and Large $N_c$ QCD \cite{Witten}, if the tree level structures 
coming from Large $N_c$ QCD for the meson-meson scattering are introduced 
in a way consistent with $\chi PT$ and then are properly unitarized 
in the way we show here. Contrary to what happens for the vector channels, 
where the tree level contributions determine the states which appear in the 
scattering, we will see that for the scalar sector the unitarization of the 
${\mathcal{O}}(p^2)$ $\chi PT$ amplitude is strong 
enough to produce meson-meson states, as for example, the $\sigma(500)$, 
$a_0(980)$, $\kappa(900)$ and a strong contribution to the $f_0(980)$. All these 
states, except the $f_0(980)$, disappear for Large $N_c$ QCD because they 
originate from effects which are subleading in $1/N_c$ counting rules 
(loops in the s-channel). We will see below that the origin for such a 
different 
behavior between the $\rho$ and the $\sigma$ will be just a numerical factor 
1/6 between the P and S-wave ${\mathcal{O}}(p^2)$ $\chi PT$ amplitude. 
Note that in a $n$-loop calculation this gives rise to a relative 
suppression factor of $1/6^{n+1}$ of the P-wave loops respect to the S-wave ones. 

\section{Formalism}

Let us consider in the first place the elastic case, corresponding to the 
scattering of two particles of masses $m_1$ and $m_2$ respectively. We will 
also allow for several coupled channels at the end of the section.

Let $\hbox{T}^I_L(s)$ be a partial wave amplitude with isospin $I$ and 
angular momentum $L$. Since we are dealing with $\pi$, $K$ and $\eta$ as 
asymptotic particles, which have zero spin, $L$ will also be the total 
spin of the partial wave. The projection in a definite angular momentum 
is given by:

\begin{equation}
\label{identical}
\hbox{T}^I_L(s)=\frac{1}{2  (\sqrt{2})^\alpha} 
\int_{-1}^{1} d\cos \theta \, \hbox{T}^I(s,cos \theta) P_L(\cos \theta)
\end{equation} 
where $(\sqrt{2})^\alpha$ is a symmetry factor to take care of the presence 
of identical particles states as $\eta \eta$ or $\pi\pi$, this last in 
the isospin limit. The index $\alpha$ can be 0,1 or 2 depending of the number
 of times
 these identical particle states appear in the corresponding partial wave
  amplitude. For instance, $\alpha=2$ for $\pi \pi \rightarrow \pi \pi$,
   $\alpha=1$ for $\eta \eta \rightarrow K \bar{K}$, $\alpha=0$ for $K \pi
    \rightarrow K \pi$ and so on. $P_L(\cos \theta)$ is the Legendre 
    polynomial of $L^{th}$ degree.

A $\hbox{T}^I_L(s)$ partial wave amplitude has two kinds of cuts. The right hand cut 
required by unitarity and the unphysical cuts from crossing 
symmetry. In our chosen normalization, the right hand cut leads to the equation 
(in the 
following discussions we omit the superindex $I$, although, it should be 
kept in mind that we always refer to a definite isospin):

\begin{eqnarray}
\label{rhc}
\hbox{Im T}_L^{-1}&=&-\rho (s)  
\end{eqnarray}
for $s>s_{threshold}\equiv s_{th}$. In the case of two particle scattering, 
the one we are concerned about, $s_{th}=(m_1+m_2)^2$ and $\rho(s)$ is 
given by: 

\begin{equation}
\label{sigma}
\rho (s)=\frac{p}{8\pi \sqrt{s}}
\end{equation}
with 
\begin{equation}
\label{p}
p=\frac{\sqrt{(s-(m_1+m_2)^2)(s-(m_1-m_2)^2)}}
{2 \sqrt{s}}\equiv \frac{\lambda^{1/2}(s,m_1^2,m_2^2)}{2 \sqrt{s}}
\end{equation}
the center mass (CM) three momentum of the two meson system.

The unphysical cuts comprise two types of cuts in the complex s-plane. For 
processes of the type $a+a\rightarrow a+a$ with $m_1=m_2=m_a$, there is only a 
left hand cut for $s<s_{Left}$. However for those ones of the type 
$a+b\rightarrow a+b$ with $m_1=m_a$ and $m_2=m_b$, apart from a left hand cut 
there is also a circular cut in the complex s-plane for $|s|=m_2^2-m_1^2$, 
where we have taken $m_2>m_1$. In the rest of this section, for simplicity 
in the formalism, we will just refer to the left hand cut as if it was the 
full unphysical cuts. This will
be enough for our purposes in this section. In any case, if we worked in the
complex $p^2$-plane all the cuts will be linear cuts and then only the left hand
cut will appear in this variable.

The left hand cut, for $s<s_{Left}$, reads:

\begin{equation}
\label{lhc}
\hbox{T}_L(s+i\epsilon)-\hbox{T}_L(s-i\epsilon)=2 i \hbox{Im T}_L(s)
\end{equation}

The standard way of solving eqs. (\ref{rhc}) and (\ref{lhc}) is the N/D 
method \cite{Chew1}. In this method a $\hbox{T}_L$(s) partial wave is 
expressed as a quotient of two functions,

\begin{equation}
\label{n/d}
\hbox{T}_L(s)=\frac{\hbox{N}_L(s)}{\hbox{D}_L(s)}
\end{equation}
with the denominator function $\hbox{D}_L(s)$, bearing the right hand cut and
the numerator function $\hbox{N}_L(s)$, the unphysical cuts.
 
 In order to take explicitly into account the behavior of a partial wave 
 amplitude near threshold, which vanishes like $p^{2L}\equiv \nu^L$, we
 consider the new quantity, $\hbox{T}'_L$, given by:

\begin{equation}
\label{T'}
\hbox{T}'_L(s)=\frac{\hbox{T}_L(s)}{\nu^L}
\end{equation}
which also satisfy relations of the type of eqs. (\ref{rhc}) and 
(\ref{lhc}). So that we can write:

\begin{equation}
\label{n/d'}
\hbox{T}'_L(s)=\frac{\hbox{N}'_L(s)}{\hbox{D}'_L(s)}
\end{equation}

From eqs. (\ref{rhc}), (\ref{lhc}) and (\ref{T'}), $\hbox{N}'_L(s)$ and 
$\hbox{D}'_L(s)$ will obey the following equations:

\begin{equation}
\label{eqs1}
\begin{array}{ll}
\hbox{Im  D}'_L=\hbox{Im T}'^{-1}_L \; \hbox{N}'_L=-\rho(s)
\hbox{N}'_L \nu^L,&  s>s_{th} \\
\hbox{Im D}'_L=0, &  s<s_{th}
\end{array}
\end{equation}
\begin{equation}
\label{eqs2}
\begin{array}{ll}
\hbox{Im  N}'_L=\hbox{Im T}'_L \; \hbox{D}'_L, & s<s_{Left}  \\
\hbox{Im N}'_L=0, & s>s_{Left}
\end{array}
\end{equation}

Since $\hbox{N}'_L$ and $\hbox{D}'_L$ can be simultaneously multiplied by any 
arbitrary real analytic function without changing its ratio, $\hbox{T}'_L$, 
nor eqs. (\ref{eqs1}) and (\ref{eqs2}), we will consider in the following that 
$\hbox{N}'_L$ is free of poles and thus, the poles of a partial wave 
amplitude will correspond to the zeros of $\hbox{D}'_L$.

Using dispersion relations for $\hbox{D}'_L(s)$ and $\hbox{N}'_L(s)$, we 
write from eqs. (\ref{eqs1}) and (\ref{eqs2}):

\begin{equation}
\label{d'}
\hbox{D}'_L(s)=-\frac{(s-s_0)^n}{\pi}\int^\infty_{s_{th}} ds' 
\frac{\nu(s')^L \rho(s') \hbox{N}'_L(s')}{(s'-s)(s'-s_0)^n}+\sum_{m=0}^{n-1}
\overline a _m s^m
\end{equation}
where $n$ is the number of subtractions needed such that 

\begin{equation}
\label{n}
\displaystyle \lim_{s \to \infty} \frac{\hbox{N}'_L(s)}{s^{n-L}}= 0
\end{equation}
since, from eq. (\ref{sigma})
\begin{equation}
\displaystyle \lim_{s \to \infty} \frac{\nu^L \rho(s)}{s^L}=\frac{1}
{4^{L+2} \pi}
\end{equation}

On the other hand, from eqs. (\ref{lhc}) and (\ref{T'}), consistently with 
eq. (\ref{n})                                       

\begin{equation}
\label{n2}
\hbox{N}'_L(s)=\frac{(s-s_0)^{n-L}}{\pi}\int_{-\infty}^{s_{Left}} ds' 
\frac{\hbox{Im  T}_L(s') \hbox{D}'_L(s')}{\nu(s')^L (s'-s_0)^{n-L} (s'-s)}+
\sum_{m=0}^{n-L-1} \overline a '_m s^m
\end{equation}                         

Eqs. (\ref{d'}) and (\ref{n2}) constitute a system of integral equations 
the input of which is given by $\hbox{Im  T}_L(s)$ along the left hand cut. 

However, eqs. (\ref{d'}) and (\ref{n2}) are not the most general solution 
to eqs. (\ref{eqs1}) and (\ref{eqs2}) because of the possible presence of zeros of $\hbox{T}_L$ 
which do not originate when solving those equations. These zeros have to be 
included explicitly and we choose to include them through poles in 
$\hbox{D}'_L$ (CDD poles after ref. \cite{Castillejo}). 
Following this last reference, let us write along the 
real axis

\begin{equation}
\label{dl1}
\hbox{Im D}'_L(s)=\frac{d \lambda (s)}{ds}
\end{equation} 

Then by eq. (\ref{eqs1}), 

\begin{equation}
\label{dl2}
\begin{array}{ll}
{\displaystyle \frac{d\lambda}{ds}}=-\rho(s) \nu^L\hbox{N}'_L, & s>s_{th} \\
 & \\
{\displaystyle \frac{d\lambda}{ds}}=0, & s<s_{th}
\end{array}
\end{equation}

Let $s_i$ be the points along the real axis where $\hbox{T}'_L(s_i)=0$. Between 
two consecutive points, $s_i$ and $s_{i+1}$, we will have from 
eq. (\ref{dl2})

\begin{equation}
\label{l1}
\lambda(s)=-\int_{s_i}^{s} \nu(s')^L \rho(s') \hbox{N}'_L(s') ds'+
\lambda(s_i)
\end{equation}
with $\lambda(s_i)$ unknown because the inverse of $\hbox{T}'_L(s_i)$ is 
not defined. Thus, we may write

\begin{equation}
\label{l2}
\lambda(s)=-\int_{s_{th}}^s \nu(s')^L \rho(s') \hbox{N}'_L(s') ds'+
\sum_i \lambda(s_i)\theta(s-s_i)
\end{equation}
with $\theta(s)$ the usual Heaviside function.

From eqs. (\ref{dl1}) and (\ref{l2}), it follows that 

\begin{eqnarray}
\hbox{D}'_L(s)&=&\frac{(s-s_0)^n}{\pi}\int_{s_{th}}^\infty \frac{
\hbox{Im  D}'_L(s') ds'}{(s'-s)(s'-s_0)^n}+\sum_{m=0}^{n-1} \overline a_m
s^m = \sum_{m=0}^{n-1} \overline a_m s^m \nonumber\\ 
&-&\frac{(s-s_0)^n}{\pi}\int_{s_{th}}^\infty \frac{\nu(s')^L \rho(s') 
\hbox{N}'_L(s')}{(s'-s)(s'-s_0)^n}ds' +
\frac{(s-s_0)^n}{\pi} \int_{s_{th}}^\infty \frac{\sum_i \lambda(s_i) 
\delta(s'-s_i)}{(s'-s)(s'-s_0)^n}ds'\nonumber\\
&=&-\frac{(s-s_0)^n}{\pi}\int_{s_{th}}^\infty \frac{\nu(s')^L \rho(s') 
\hbox{N}'_L(s')}{(s'-s)(s'-s_0)^n}ds'+\sum_{m=0}^{n-1} \overline a_m s^m
\nonumber\\ &+&
\sum_i \frac{\lambda(s_i)}{\pi (s_i-s_0)^n}\frac{(s-s_0)^n}{s_i-s}
\label{tower1}
\end{eqnarray}

  Eq. (\ref{tower1}) can also  be obtained from eq. (\ref{eqs1}) and 
  use of the Cauchy theorem for complex integration and allowing for the 
presence of poles of $\hbox{D}'_L$ (zeros of $\hbox{T}'_L$) inside and along 
the integration contour, 
which is given by a circle in the infinity deformed to engulf the real axis
along the right hand cut, $s_{th}<s'<\infty$. In this way one can also consider 
the possibility of there being higher order zeros and that some of the $s_i$ 
could have non-vanishing imaginary part (because of the Schwartz theorem, 
$s_i^*$ will be another zero of $\hbox{T}'_L(s)$). However, as we will see
 below for $L\leq 1$, when considering chiral 
symmetry in the Large $N_c$ QCD limit, the zeros will appear on the real 
axis and also as simple zeros. In general, using $\hbox{T}'_L$ instead 
of $\hbox{T}_L$, we avoid working with $L^{th}$ order poles 
of $\hbox{D}_L$ at threshold in the dispersion relation given by eq.(\ref{tower1}).
 
The last term in the right hand side of eq. (\ref{tower1}) can also be 
written in a more convenient way avoiding the presence of the 
subtraction point $s_0$. To see this, note that 

\begin{eqnarray}
\label{cano}
\frac{(s-s_0)^n}{s-s_i}&=&(s-s_0)^{n-1} \frac{s-s_i+s_i-s_0}{s-s_i}\nonumber
\\
&=&(s-s_0)^{n-1}(1+\frac{s_i-s_0}{s-s_i})=(s-s_0)^{n-1}+(s_i-s_0)\frac
{(s-s_0)^{n-1}}{s-s_i} \nonumber\\
&=&\sum_{i=0}^{n-1}(s-s_0)^{n-1-i}(s_i-s_0)^i+\frac{(s_i-s_0)^n}{s-s_i}
\end{eqnarray}

The terms 
\begin{displaymath}
\sum_{i=0}^{n-1}(s-s_0)^{n-1-i}(s_i-s_0)^i 
\end{displaymath}
can be reabsorbed in 
\begin{displaymath}
\sum_{m=0}^{n-1} \overline a _m s^m
\end{displaymath}

As a result we can write  

\begin{equation}
\label{d'2}
\hbox{D}'_L(s)=-\frac{(s-s_0)^n}{\pi}\int_{s_{th}}^\infty 
\frac{\nu(s')^L \rho(s') 
\hbox{N}'_L(s')}{(s'-s)(s'-s_0)^n}+\sum_{m=0}^{n-1} \widetilde{a}_m 
s^m+\sum_i \frac{\widetilde{\gamma}_i}{s-s_i}
\end{equation} 
with $\widetilde{a}_m$ ($n-1\geq m \geq 0$) and $\widetilde{\gamma}_i$, $s_i$ 
($i\geq 0$), arbitrary parameters. However, if some of the $s_i$ is complex 
there will be another $s_j$ such that $s_j$=$s_i^*$ and $\widetilde{\gamma}_j=
\widetilde{\gamma}_i^*$, as we explained above. Each term of the last 
sum in eq. (\ref{d'2}) is referred as a CDD pole after \cite{Castillejo}.

 Eqs. (\ref{d'2}) and (\ref{n2}) stand for the general integral equations 
 for $\hbox{D}'_L$ and $\hbox{N}'_L$, respectively. Next we make the 
 approximation of neglecting the left hand cut, that is, we set $\hbox{Im 
 T}_L(s)=0$ in eq. (\ref{n2}). Thus one has:

\begin{equation}
\label{naprox}
\hbox{N}'_L(s)=\sum_{m=0}^{n-L-1} \widetilde a '_m s^m
\end{equation} 

As a result, $\hbox{N}'_L(s)$ is just a polynomial of degree $\leq n-L-1$
\footnote{One can always make that $n\geq 
L+1$ just by multiplying $\hbox{N}'_L$ and $\hbox{D}'_L$ by 
$\displaystyle{s^k}$ with $k$ large enough.} . So we 
can write, 

\begin{equation}
\label{ojo}
\hbox{N}'_L(s)={\mathcal{C}} \prod_{j=1}^{n-L-1}(s-s_j)
\end{equation}

In eq. (\ref{ojo}) it is understood that if $n-L-1$ is zero $\hbox{N}'_L$ is
just a constant. Thus, the only effect of $\hbox{N}'_L$ will 
be, apart of the normalization constant ${\mathcal{C}}$, the inclusion, 
at most, of $n-L-1$ zeros to $\hbox{T}'_L(s)$. But we can always divide 
$\hbox{N}'_L$ and $\hbox{D}'_L$ by eq. (\ref{ojo}). The net result is that, 
when the left hand cut is neglected, it is always possible to take 
$\hbox{N}'_L(s)=1$ 
and all the zeros of $\hbox{T}'_L(s)$ will be CDD poles of the 
denominator function. In this way,

\begin{eqnarray}
\label{fin/d}
\hbox{T}'_L(s)&=&\frac{1}{\hbox{D}'_L(s)}\nonumber\\
\hbox{N}'_L(s)&=&1\nonumber\\
\hbox{D}'_L(s)&=&-\frac{(s-s_0)^{L+1}}{\pi}\int_{s_{th}}^\infty ds' \frac
{\nu(s')^L \rho (s')}{(s'-s)(s'-s_0)^{L+1}}+\sum_{m=0}^L a_m s^m+
\sum_i^{M_L} \frac{R_i}{s-s_i}
\end{eqnarray}
  
The number of free parameters present in eq. (\ref{fin/d}) is $L+1+2\varrho$, 
where $\varrho$ is the number of CDD poles, $M_L$, minus the number of complex 
conjugate pairs of $s_i$. These free parameters have a clear physical
 interpretation. Consider first the term $2 \varrho$ which comes from the 
 presence of CDD poles in $\hbox{D}'_L(s)$, eq. (\ref{fin/d}). In \cite{Chew2} 
 the presence of CDD poles was linked 
to the possibility of there being elementary particles with the same 
quantum numbers as those of the partial wave amplitude, that is, particles 
which are not originated from a given `potential' or exchange forces between 
the scattering states. One can think that given a $\hbox{D}'_L(s)$ we can 
add a CDD pole and adjust its two parameters in order to get a zero of the real 
part of the new 
$\hbox{D}'_L(s)$ with the right position and residue, having a resonance/bound 
state with the desired mass and coupling. 
In this way, the arbitrary parameters that come with a CDD pole can be 
related with the coupling constant and mass of the resulting particle. This is 
one possible interpretation of the presence of CDD poles. However, as we are 
going to see below, these poles can also enter just to ensure the presence of 
zeros required by the underlying theory, in this case QCD, as the Adler zeros 
for the S-wave 
meson-meson interaction. The derivative of the partial wave amplitude at the 
zero will fix the other CDD parameter, $\widetilde{\gamma}_i$. With respect 
to the contribution $L+1$ to the number of free 
parameters coming from the angular momentum $L$, it appears just because we 
have explicitly established the behavior of a partial wave amplitude 
close to threshold, vanishing as $\nu ^L$. This is required by the 
centrifugal barrier effect, well known from Quantum Mechanics.

It should be stressed that eq. (\ref{fin/d}) is the most general structure 
that an elastic partial wave amplitude, with arbitrary L, has when the left 
hand cut is neglected. The free parameters that appear there are fitted to 
the experiment or calculated from the basic underlying theory. In our case the 
basic dynamics is expected to be QCD, but eq. (\ref{fin/d}) could also be 
applied to other scenarios beyond QCD as the Electroweak Symmetry Breaking 
Sector (which also has the symmetries \cite{ESBS} used to derived eq. 
(\ref{fin/d}), as far as it is known).

Let us come back to QCD and split the subtraction constants $a_m$ of 
eq. (\ref{fin/d}) in two pieces

\begin{equation}
\label{split}
a_m=a^L_m+a^{SL}_m(s_0)
\end{equation}
The term $a_m^L$ will go as $N_c$, because in the $N_c \rightarrow \infty$ limit, 
the meson-meson amplitudes go as $N_c^{-1}$ \cite{Witten}. Since 
the integral in eq. (\ref{fin/d}) is $\mathcal{O}(1)$ in this counting, the 
subleading term $a_m^{SL}(s_0)$ is of the same order and depends on the 
subtraction point $s_0$. This implies that eq.
 (\ref{fin/d}), when $N_c \rightarrow 
\infty$, will become 

\begin{equation}
\label{limitd'}
\hbox{D}'_L(s)\equiv \hbox{D}'^{\infty}_L(s)=\sum_{m=0}^L 
a^L_m s^m+\sum_i^{M^\infty_L}\frac{R^\infty_i}{s-s_i}
\end{equation}    
where $R^\infty_i$ is the leading part of $R_i$ and $M^\infty_L$ counts the
number of leading CDD poles.

Clearly eq. (\ref{limitd'}) represents tree level structures, contact and pole 
terms,  which have nothing to do with any kind of potential scattering, 
which in Large $N_c$ QCD is suppressed. 

In order to determine eq. (\ref{limitd'}) we will make use of $\chi PT$ 
\cite{Retaila} and of the paper \cite{EP}. In this latter work it is shown 
the way to include resonances with spin $\leq 1$ consistent with chiral 
symmetry at lowest order in the chiral power counting. It is also seen 
that, when integrating out the resonance fields, the contributions of the 
exchange of these resonances essentially saturate the next to leading 
$\chi PT$ Lagrangian. We will make use of this result in order to state 
that in the inverse of eq. (\ref{limitd'}) the contact terms come just from 
the lowest order $\chi PT$ Lagrangian and the pole terms from the exchange 
of resonances in the s-channel in the way given by \cite{EP} (consistently 
with our approximation of neglecting the left hand cut the exchange 
of resonances in crossed channels is not considered). In the latter statement 
it is assumed that the result 
of \cite{EP} at ${\mathcal{O}}(p^4)$ is also applicable to higher orders. That is, 
local terms appearing in $\chi PT$ and from eq. (\ref{limitd'}) of order higher than 
${\mathcal{O}}(p^4)$ are also saturated from the exchange of resonances for 
$N_c>>1$, where loops are suppressed.

Let us prove that eq. (\ref{limitd'}) can accommodate the tree level amplitudes 
coming from lowest order $\chi PT$ \cite{Retaila} and 
the Lagrangian given in $\cite{EP}$ for the coupling of resonances (with spin 
$\leq 1$) with the lightest pseudoscalars ($\pi$, $K$ 
and $\eta$).

Following \cite{EP}, one can write the exchange of a resonance (with 
spin $\leq$ 1) divided by $\nu^L$ as:

\begin{eqnarray}
\label{ep1}
\begin{array}{ll}
{\displaystyle \frac{(c_i s +c'_i m^2)^2}{M_i^2-s}} &\; ; \; L=0 
\end{array} \nonumber\\
\begin{array}{ll}
{\displaystyle \frac{d_i s}{M_i^2-s}} &\; ; \; L=1
\end{array}
\end{eqnarray}
where $M_i$ is the mass of the $i^{th}$ resonance, $m^2$ is some 
combination of squared masses of the lightest pseudoscalars and $c_i$, 
$c'_i$ and $d_i$ are arbitrary constants with $d_i \geq 0$. 

The lowest order $\chi PT$ partial wave amplitude ($L \leq 1$), divided 
again by $\nu^L$, can be written schematically as:

\begin{eqnarray}
\label{ep2}
\begin{array}{ll}
as+a' \hat{m}^2 &\; ; \; L=0\\
\end{array}\nonumber\\
\begin{array}{ll}
b & \; ; \; L=1
\end{array}
\end{eqnarray}
with $\hat{m}^2$ another combination of squared masses of the lightest 
pseudoscalars.

Thus, in Large $N_c$ QCD, the partial wave amplitudes will have the following 
structure, after omitting the exchange of resonances 
in crossed channels and contact terms of order higher than $\mathcal{O}(p^2)$

\begin{eqnarray}
\label{t'1}
\hbox{T}'_0(s)\equiv \hbox{T}'^{ \infty}_0(s)=as+a'\hat{m}^2+\sum_{i=1}^{R_0} 
\frac{(c_i s+c'_i m^2)^2}{M_i^2-s}\nonumber \\
\hbox{T}'_1(s)\equiv\hbox{T}'^{\infty}_1(s)=b+\sum_{i=1}^{R_1} 
\frac{d_i s}{M_i^2-s}
\end{eqnarray} 

In the former equation it is understood that if $R_L=0$ the sum does not 
appear. 

Let us now study the inverse of eq. (\ref{t'1}) in order to connect with $
\hbox{D}'^{\infty}_L$. With $\xi_L$, the number of zeros of 
$\hbox{T}'^{ \infty}_L$, we can write:

\begin{equation}
\label{t'2}
\frac{1}{\hbox{T}'^{\infty}_L(s)}={\mathcal{A}}_L \frac{\prod_{i=1}^{R_{L}} 
(s-M_i^2)}{\prod_{r=1}^{\xi_{L}}(s-s_r)}
\end{equation} 

In the former equation, if $R_L=0$ or $\xi_L=0$, the corresponding product 
must be substituted by 1. ${\mathcal{A}}_L$ is just a constant. Note that from
eqs. (\ref{t'1}) $\xi_0\leq R_0+1$ and $\xi_1\leq R_1$, from simple counting of
the degree of the polynomials that appear in the numerators after writing eqs.
(\ref{t'1}) as rational functions. 

On the other hand, note from 
eqs. (\ref{ep1}) that, for $s\geq 0$, the amplitude from the exchange of 
resonances, both for 
$L=0$ and 1, is positive below the mass of the resonance and negative above 
it. This implies that in the interval, $M_{i+1}^2\geq s \geq M_i^2$, by 
continuity, there will be a zero of $\hbox{T}'^{ \infty}_L$. In this way, 

\begin{equation}
\label{ceros2}
\xi_L\geq R_L-1
\end{equation}

For $L=0$, apart from the zeros between resonances, one has also the 
requirement from chiral symmetry of the presence of the Adler zero, 
along the real axis and below threshold. Thus, $\xi_0\geq R_0$ and we can
write

\begin{equation}
\label{ceros3}
R_L-L+1 \geq \xi_L \geq R_L-L
\end{equation}

Note that since the $R_L-L$ zeros of $\hbox{T}^\infty_L$ are real, the possible
single additional zero from the upper limit of eq. (\ref{ceros3}) must be also
real, since a complex one would imply its conjugate too. This is so because all
the coefficients in eq. (\ref{t'1}) are real.

Let us do the counting of zeros and resonances from eq. (\ref{limitd'}). The
number of zeros of $\hbox{T}'^{ \infty}_L$ is equal to the number of CDD
poles of $\hbox{D}'^\infty_L$, $M^\infty_L$. Hence 

\begin{equation}
\label{pepe1}
\xi_L=M_L^\infty
\end{equation}

On the other hand the number of poles of $\hbox{T}'^{\infty}_L$ is equal to
the number of zeros of $\hbox{D}'^\infty_L$ which has the following upper 
limit

\begin{equation}
\label{pepe2}
R_L \leq M_L^\infty+L \equiv \xi_L+L
\end{equation} 

Thus,

\begin{equation}
\label{pepe3}
\xi_L\geq R_L-L
\end{equation}
which gives the same lower limit as in eq. (\ref{ceros3}). Let us recall that
the zeros of eq. (\ref{t'2}) are real, as discussed above. This, together with
the requirement of $\hbox{T}'^{\infty}_L(s)$ being a real function on the
real axis, forces all parameters in eq. (\ref{limitd'}) to be real. Then the
number of free parameters in eq. (\ref{limitd'}) are $2 M_L^\infty+L+1=2
\xi_L+L+1$. When fixing the $s_i$ parameters in eq. (\ref{limitd'}) to the $s_r$
parameters in eq. (\ref{t'2}) this reduces in $\xi_L$ the number of free
parameters in eq. (\ref{limitd'}). By fixing the arbitrary constant $\mathcal{A}_L$
of eq. (\ref{t'2}) this leaves us with $\xi_L+L$ free parameters in eq.
(\ref{limitd'}). Imposing now the position of the $R_L$ resonances of eq.
(\ref{t'2}) to be at $M_i^2$ we have $R_L$ additional constraints. Hence we
should have

\begin{equation}
\label{pepe4}
\xi_L+L-R_L\geq 0
\end{equation}
which actually holds, as seen above in eq. (\ref{pepe3}). As a consequence eqs.
(\ref{t'1}) can always be cast in the form of eq. (\ref{limitd'})

Let us define the function $g_L(s)$ by 

\begin{equation}
\label{g(s)}
g_L(s) \nu^L=\sum_{m=0}^L a^{SL}_m(s_0) s^m-\frac{(s-s_0)^{L+1}}{\pi}
\int_{s_{th}}^\infty ds' \frac{\nu(s')^L \rho(s')}{(s'-s)(s'-s_0)^{L+1}}
\end{equation}

and using the notation

\begin{equation}
\label{Tinfty}
\hbox{T}^\infty_L(s)=\nu^L \big[ \sum_{m=0}^L a^L_m s^m
+\sum_{i=1}^{M_L} \frac{R_i}{s-s_i} \big]^{-1}
\end{equation}

results that from eq. (\ref{fin/d}) one has

\begin{equation}
\label{TL}
\hbox{T}_L(s)=\big[ 1/\hbox{T}^\infty_L(s)+g_L(s) \big]^{-1}
\end{equation}

The physical meaning of Eq. (\ref{TL}) is clear. The $\hbox{T}_L^\infty$ 
amplitudes correspond to the tree levels structures present before
unitarization. The unitarization is then accomplished through 
the function $g_L(s)$. It is interesting at this point to connect with 
the most popular K-matrix formalism to obtain unitarized amplitudes. In this 
case one writes:

\begin{equation}
\label{Kmatrix}
T_L(s)=[K_L(s)^{-1}-i\rho (s)]^{-1}
\end{equation}
we see that the former equation is analogous to eq. (\ref{TL}) with 
$\hbox{T}^{\infty \, -1}_L(s)+\hbox{Re}\, g_L(s) =  K_L(s)^{-1}$. 

From the former comments it should be obvious the generalization of eq.
(\ref{TL}) to coupled channels. In this case, 
$\hbox{T}^\infty_L(s)$ is a matrix determined by the 
tree level partial wave amplitudes given by the lowest order 
$\chi PT$ Lagrangian \cite{Retaila} and the exchange of resonances \cite{EP}. 
For instance, $[T^\infty_L(s)]_{11}=T^{(2)}_{11}+T^R_{11}$, 
$[T^\infty_L(s)]_{12}=T^{(2)}_{12}+T^R_{12}$ and so on. Where $T^{(2)}$ is 
the matrix of the ${\mathcal{O}}(p^2)$, 
$\chi PT$ partial wave amplitudes \cite{Retaila} and $T^R$ the corresponding 
one from the exchange of resonances \cite{EP} in the s-channel. Once we have 
$\hbox{T}^\infty_L(s)$ its inverse is the one 
which enters in eq. (\ref{TL}). Because $\hbox{N}'_L(s)$ is proportional to 
the  identity, $g_L(s)$ will be a diagonal matrix, accounting for the right
 hand cut, as in the elastic case. In this way, unitarity, which 
 in coupled channels reads (above the thresholds of the channels $i$ and $j$) 

\begin{equation}
\label{ccu}
[\hbox{Im  T}^{-1}_{L}]_{ij}=-\rho_{ii}(s) \delta_{ij}
=\hbox{Im  g}_L(s)_{ii} \delta_{ij}
\end{equation}
is fulfilled. The matrix element $g_L(s)_{ii}$ obeys eq. (\ref{g(s)}) with the 
right masses corresponding to the channel $i$ and its own subtraction constants 
$a^{SL}_i(s_0)$.

In the present work the coupling constants and resonance masses contained in 
$\hbox{T}^\infty_L(s)$ are fitted to the experiment. The same 
happens with the $a^{SL}_i$ although, as we will discuss below, they are 
related by $SU(3)$ considerations.

In Appendix A the already stated coupled channel version of eq. (\ref{TL}) is 
deduced directly from the N/D method in coupled channels \cite{Bjorken}. 
       
\section{The ideal case: elastic vector channels.}

In this section we are going to study the $\pi\pi$ and $K\pi$ scattering 
with I=$L$=1 and  I=1/2,$L$=1, respectively. It will be shown, as mentioned in the 
introduction, that these reactions can be understood just by invoking:

$$
\begin{array}{ll}
1) & \hbox{Chiral Symmetry} \\
2) & \hbox{Large}\; N_c \; \hbox{QCD}\\
3) & \hbox{Unitarity} 
\end{array}
$$

A priori, one could think that relations coming from the $N_c\rightarrow 
\infty$ limit should work rather accurately at the phenomenological levels in 
these channels due to the predominant role of the $\rho$ and $K^*$ poles 
over subleading effects in 1/$N_c$, as unitarity loops.

In eq. (\ref{fin/d}) the $L$=1 zero at threshold is included in the N$_1$
function, eq. (\ref{n/d}), which can be taken as

\begin{eqnarray}
\label{porc}
\hbox{N}_1&=&\nu \, \hbox{N}'_1=\nu \\ \nonumber
\hbox{D}_1&=&\hbox{D}'_1 
\end{eqnarray}
with N$_1'$ and D$_1'$ defined in eq. (\ref{n/d'}) and given in eq. 
(\ref{fin/d}). However, in this chapter we will also include the P-wave
threshold factor as a CDD pole in the D$_1$ function. In this way, we will treat
the S-wave Adler zero and the P-wave threshold zero in the same footing, which will
make the comparision among both partial waves more straightforward. The final
result can be derived in the same way as eq. (\ref{fin/d}) but working
directly with T$_1$ instead of T$_1'$. Thus:

\begin{eqnarray}
\label{d1}
\hbox{N}_1(s)&=&1 \\ \nonumber
\hbox{D}_1(s)&=&\sum_{i}\frac{\gamma_i}{s-s_i}+a-\frac{s-s_0}{\pi} 
\int_{s_{th}}^\infty ds' \frac{\rho(s')}{(s'-s)(s'-s_0)}
\end{eqnarray}

Of course, the partial wave amplitude T$_1$ will be the same than before since
we have just divided at the same time N$_1$ and D$_1$ by $\nu=p^2$. As a simple
and explicit example that eq. (\ref{fin/d}) together with eq. (\ref{porc}) are
equivalent to eq. (\ref{d1}), let us consider the scattering of two pions. 
After multiplying N$_1$ and D$_1$ in eq. (\ref{porc}) by 4, we divide
both functions by $4\,\nu=s-s_A$, with $s_A=4\,m_\pi^2$. Keeping in mind eq.
(\ref{fin/d}), one still has a sum over the former CDD poles together with 
another one at $s=s_A$. The subtraction polynomial of order one in
$s$ transforms into a CDD pole at $s=s_A$ plus a constant. Finally, the dispersive
integral gives rise to that of the former equation (\ref{d1}) together with a
constant plus a CDD pole at $s=s_A$ --- to see this just add and subtract $s_0$
 in $4\,\nu(s')$ in the integral of eq. (\ref{fin/d}). Hence, the structure 
 given in eq. (\ref{d1}) is obtained.

\vspace{-3.cm}
\begin{figure}[H]
\centerline{
\protect
\hbox{
\psfig{file=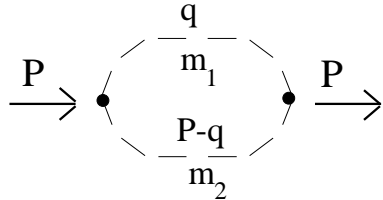,width=0.4\textwidth,angle=-90}}}
\caption{Loop giving rise to the $g_0$ function.}
\end{figure}

The integral in eq. (\ref{d1}) will be evaluated making use of dimensional 
regularization. It can be identified up to a constant to the loop represented 
in Fig. 1. This identification is consequence of the fact that both the integral
in eq. (\ref{d1}) and the loop given in Fig. 1 have the same cut and the same
imaginary part along this cut, as it can be easily checked.

Following eq. (\ref{g(s)}) we define

\begin{eqnarray}
\label{g(s)dr}
g_0(s)&=&a^{SL}(s_0)-\frac{s-s_0}{\pi}\int_{s_{th}}^\infty ds' 
\frac{\rho(s')}{(s'-s)(s'-s_0)}\nonumber\\
&=&\frac{1}{(4\pi)^2}\Bigg[\tilde{a}^{SL}(\mu)+\log \frac{m_2^2}{\mu^2} 
-\frac{m_1^2-m_2^2+s}{2s}\log\frac{m_2^2}{m_1^2}-\frac{\lambda^{1/2}(s,m_1^2,
m_2^2)}{2s} \cdot \nonumber \\
&& \cdot  \log\Big( \frac
{m_1^2+m_2^2-s+\lambda^{1/2}(s,m_1^2,m_2^2)}
{m_1^2+m_2^2-s-\lambda^{1/2}(s,m_1^2,m_2^2)}\bigg)\Bigg]
\end{eqnarray}  
for $s\geq s_{th}$. For $s<s_{th}$ or $s$ complex one has the analytic 
continuation of eq. (\ref{g(s)dr}). The function $\lambda^{1/2}(s,m_1^2,
m_2^2)$ was already introduced in eq. (\ref{p}). The regularization 
scale $\mu$, appearing in the last formula of eq. (\ref{g(s)dr}), plays 
a similar role than the arbitrary subtraction point $s_0$ in the first formula 
of eq. (\ref{g(s)dr}).
This similarity is consequence of the fact that both $\mu$ and $s_0$ can have 
any arbitrary value but the resulting function $g_0(s)$ is independent of this 
particular value because of the change in the subtraction constant, 
$\tilde{a}^{SL}(\mu)$ for dimensional regularization or $a^{SL}(s_0)$ for the 
dispersion integral. The $\tilde{a}^{SL}(\mu)$ `constant' will change under a 
variation of the scale $\mu$ to other one $\mu'$ as 

\begin{equation}
\label{scale}
\tilde{a}^{SL}(\mu')=\tilde{a}^{SL}(\mu)+\log\frac{\mu'^2}
{\mu^2}
\end{equation}
in order to have $g_0(s)$ invariant under changes of the regularization scale. 
We will take $\mu=M_{\rho}=770$ MeV \cite{PDG}. The function $g_0(s)$ is 
also symmetric under the exchange $m_1\leftrightarrow m_2$ and for the 
equal mass limit it reduces to 

\begin{equation}
\label{g(s)dr2}
g_0(s)=\frac{1}{(4\pi)^2}\bigg[\tilde{a}^{SL}(\mu)+\log\frac{m_1^2}{\mu^2}
+\sigma(s)\log\frac{\sigma(s)+1}{\sigma(s)-1}\bigg]
\end{equation}
with 
\begin{equation}
\label{sigma2}
\sigma(s)=\sqrt{1-\frac{4m_1^2}{s}}
\end{equation}

After this preamble, let us consider in first place the $\pi\pi$ scattering 
with I=$L$=1. As can be seen from Fig. 2 this process is dominated by the 
$\rho$ exchange. From eq. (\ref{d1}) one has

\begin{equation}
\label{rho}
\hbox{T}^{\pi\pi}_1(s)=\bigg[\frac{\gamma^{\pi\pi}_1}{s-4m_{\pi}^2}+
\tilde{a}^L_{\pi\pi}+g^{\pi\pi}_0(s)+\sum_{i=2}\frac{\gamma_i}{s-s_i}\bigg]^{-1}
\end{equation}

The first term in the R.H.S of the last equation fixes the zero at threshold 
for a P-wave amplitude.

The tree level part of $T^{\pi\pi}_1(s)$ from ref. \cite{EP} and lowest order 
$\chi PT$ \cite{Retaila}, in the way explained in the last section, is given by: 

\begin{equation}
\label{t1n2}
\hbox{T}^{\pi\pi \, \infty}_1(s)=\frac{2}{3}\frac{p_{\pi\pi}^2}
{f^2}+g_v^2 \frac{2}{3}\frac{p_{\pi\pi}^2}{f^2}\frac{s}{M_\rho^2-s}
\end{equation} 
with $p_{\pi\pi}^2$ the three-momentum squared of the pions in the CM, 
$f=87.3$ MeV the pion decay constant in the chiral limit \cite{Retaila}. 
The deviation of $g_v^2$ with respect to unity measures the variation of the 
value of the $\rho$ coupling to two pions with respect to the KSFR relation 
\cite{KSFR}, $g_v^2=1$. In \cite{EP2} this KSFR relation is justified 
making use of Large $N_c$ QCD (neglecting loop contributions) and an 
unsubtracted dispersion relation for the pion electromagnetic form factor 
(a QCD inspired high-energy behavior).

Comparing eqs. (\ref{rho}) and (\ref{t1n2}), one need only one additional CDD
pole apart from the one at threshold and we obtain

\begin{eqnarray}
\label{aLpi}
\tilde{a}^L&=& 0 \nonumber \\
\gamma^{\pi\pi}_1&=&\frac{6 f^2 (M_\rho^2-4m_\pi^2)}
{(M_\rho^2-4m_\pi^2(1-g_v^2))}\nonumber\\
\gamma^{\pi\pi}_2&=&\frac{6 f^2}{1-g_v^2}\; \frac{g_v^2 M_\rho^2}{M_\rho^2-
(1-g_v^2)4 m_\pi^2} \nonumber \\
s_2&=&\frac{M_\rho^2}{1-g_v^2}
\end{eqnarray}

Thus, we can write our final formula for the isovector $\pi\pi$ scattering in 
the following way

\begin{equation}
\label{ftpi}
\hbox{T}^{\pi\pi}_1(s)=\bigg[\frac{\gamma^{\pi\pi}_1}{s-4m_\pi^2}+
\frac{\gamma^{\pi\pi}_2}{s-s_2}+g^{\pi\pi}_0(s) \bigg]^{-1}
\end{equation}
in terms of the parameters $g_v^2$ and $\tilde{a}^{SL}(\mu)$. Since $g_v^2$ 
is expected to be close to unity as discussed above, it is useful to consider 
the limit when $g_v^2 \rightarrow 1$, in which case $s_2 \rightarrow \infty$ 
such that

\begin{equation}
\label{5.1}
\frac{\gamma_2^{\pi\pi}}{s-s_2} \rightarrow -\frac{6 f^2}{M_\rho^2}\equiv
\tilde{a}'^{L}_{\pi\pi}
\end{equation}
in which case the second CDD pole in eq. (\ref{ftpi}), at $s_2$, moves to
infinity and the CDD pole contribution gives rise to a constant term,
$\tilde{a}'^{L}_{\pi\pi}$. In this limit we can write 

\begin{equation}
\label{5.2}
\hbox{T}^{\pi\pi}_1(s)=\big[ \frac{\gamma_1^{\pi\pi}}{s-4m_\pi^2}+
\tilde{a}'^{L}_{\pi\pi}+g^{\pi\pi}_0(s) \big]^{-1}
\end{equation}
 
Our calculated phase shifts for the vector $\pi\pi$ scattering are represented 
in the dashed line of Fig. 2, when $g_v^2$ is taken equal to one and $\tilde{a}^{SL}=0$ is set 
to zero. The agreement with the experimental data is rather good. If we had 
just taken the imaginary part of 
$g^{\pi\pi}_0(s)$ in the former equation we would have obtained basically the 
same curve. This last case corresponds just to the Breit Wigner amplitude 

\begin{eqnarray}
\label{frho}
&&\frac{2}{3}\frac{p_{\pi\pi}^2}{f^2}\; \frac{M_\rho^2}{M_\rho^2-s-\hbox{i} M_\rho
\Gamma_\rho(s)}\nonumber \\
&&\Gamma_\rho(s)=\frac{p_{\pi\pi}^3}{12 \pi f^2}\frac{M_\rho}{\sqrt{s}}
\end{eqnarray}

\begin{figure}[ht]
\centerline{
\protect
\hbox{
\psfig{file=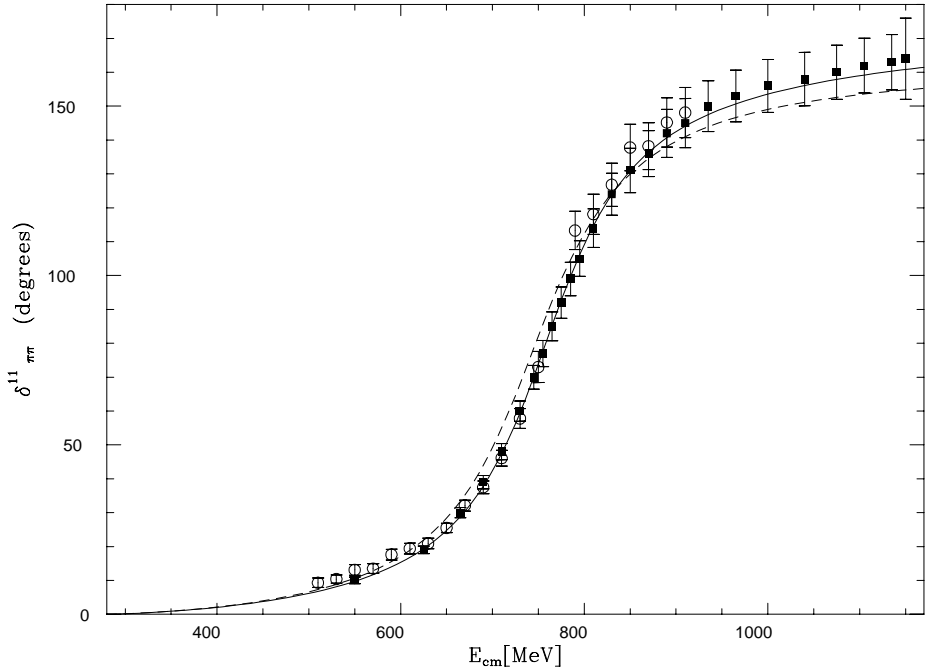,width=0.6\textwidth,angle=-90}}}
\caption{Isovector $\pi\pi$ elastic phase shifts from threshold up to 
$\sqrt{s}\leq 1.2$ GeV. The dashed line corresponds to take $g_v^2=1$ and 
$\tilde{a}^{SL}=0$. The continuum line corresponds to the simultaneous fit to 
the $\rho$ and $K^*$ channels, given by eq. (\ref{fit}). Data: circles 
\cite{29o}, squares \cite{30o}}
\end{figure}

For the I=1/2, $L$=1 $K\pi$ scattering, the tree level amplitude 
$\hbox{T}_1^{K\pi \, \infty}$, is just given 
by multiplying eq. (\ref{t1n2}) by 3/4 and substituting $p_{\pi\pi}^2$ by 
$p_{K\pi}^2=\displaystyle{\frac{\lambda(s,m_\pi^2,m_K^2)}{4s}}$ and $M_\rho$ 
by $M_K^*$=896 MeV \cite{PDG}, the mass of the neutral $K^*(890)$. Now $m_K$ is 
the mass of the kaon, 495.7 MeV \cite{PDG}. On the other hand, since 
$m_K\neq m_\pi$ instead of having just a single pole in the denominator 
function, as in eq. (\ref{rho}), for the zero at threshold, one has two simple 
poles

\begin{equation}
\label{twopo}
\frac{\gamma^{K\pi}_1}{4 p_{K\pi}^2}=\frac{\gamma^{K\pi}_1}{s_+-s_-} 
\bigg[ \frac{s_+}{s-s_+} - \frac{s_-}{s-s_-} \bigg]
\end{equation} 
with $s_+=(m_K+m_\pi)^2$ and $s_-=(m_K-m_\pi)^2$. That is, we will 
have two CDD poles but both entering with just one parameter because 
the behavior at threshold is proportional to $p_{K\pi}^2$. In our notation

$$
\lim_{s \to s_{th}}\frac{\hbox{T}^{K\pi}_1(s)}{p_{K\pi}^2}=\frac{4}
{\gamma^{K\pi}_1}
$$

For the simple and realistic case, $g_v^2=1$, we have, analogously to the case
of the $\rho$, 

\begin{equation}
\label{6.1}
\hbox{T}^{K\pi}_1(s)=\big[ \frac{\gamma_1^{K\pi}}{4 p^2_{K\pi}}+
\tilde{a}'^{L}_{K\pi}+g^{K\pi}_0(s) \big]^{-1}
\end{equation}
with 

\begin{eqnarray}
\label{aLkpi}
\gamma^{K\pi}_1 &=& \frac{8 f^2 (M_{K^*}^2-(m_K+m_\pi)^2)}
{M_{K^*}^2}\nonumber\\
\tilde{a}^L_{K\pi} &=& -\frac{8f^2}{M_{K^*}^2}
\end{eqnarray}

In the general case when $g_v^2\neq 1$, one proceeds in the same way as for the
$\rho$ introducing an extra CDD, but we shall omit the details here and the
evaluations are done directly using the final formula 

\begin{equation}
\label{7.1}
\hbox{T}_1^{K\pi}=\big[ \frac{1}{1/T_1^{K\pi \, \infty}+g^{K\pi}_0(s)} \big]^{-1}
\end{equation} 
with $T_1^{K\pi \, \infty}$ evaluated as mentioned above.

In Fig. 3, the calculated phase shifts for the I=1/2, P-wave $K\pi$ scattering 
are shown in the dashed curve for $g_v^2=1$ and $\tilde{a}^{SL}=0$. The same 
remarks, as done before for the $\rho$ when commenting Fig. 2, are 
also valid for the $K^*$. It is worth 
stressing that the dashed lines of Fig. 2 and 3 have no free parameters at all, 
and depend only on $f$ and the masses of the resonances $K^*$ and $\rho$.

\begin{figure}[ht]
\centerline{
\protect
\hbox{
\psfig{file=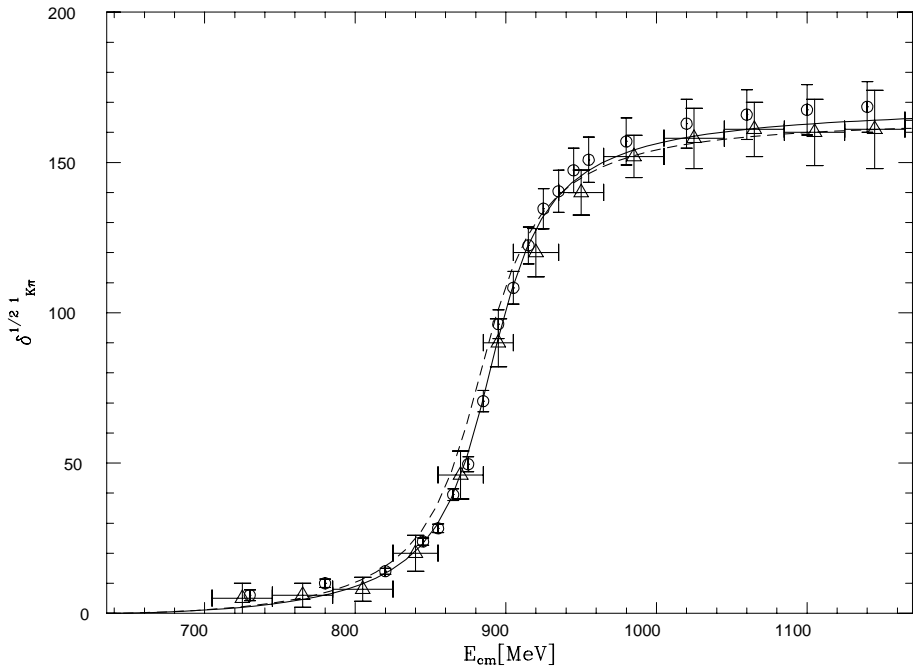,width=0.6\textwidth,angle=-90}}}
\caption{P-wave elastic $K\pi$ phase shifts with I=1/2, from threshold up to 
$\sqrt{s}\leq 1.2$ GeV.The dashed line corresponds to take $g_v^2=1$ and 
$\tilde{a}^{SL}=0$. The continuum line corresponds to the simultaneous fit 
to the $\rho$ and $K^*$ channels, given by eq. (\ref{fit}). Data: triangles 
\cite{36o}, circles \cite{390}}
\end{figure}   

The subleading constant $\tilde{a}^{SL}$ presents in $g_0(s)$, eq. 
(\ref{g(s)dr}), should be the same for the $\pi\pi$ and $K\pi$ states 
because the dependence of the loop represented in Fig. 1 
on the masses of the intermediate particles is given by eq. (\ref{g(s)dr}). 
This point can be used in the opposite sense. That is, if it is not possible 
to obtain a reasonable good fit after setting $\tilde{a}^{SL}$ to be the same 
in both channels, some kind of SU(3) breaking is missing.

 From eqs. (\ref{ftpi}) and (\ref{7.1}) we do a simultaneous 
fit to the experimental P-wave $\pi\pi$ (I=1) and 
$K\pi$ (I=1/2) phase shifts with $g_v^2$ and $\tilde{a}^{SL}$ as free 
parameters using the minimization program MINUIT. In order to make the data 
from different experiments consistent 
between each other, a systematic relative error of a $5\%$ is given to each 
data point if its own error is smaller than this bound. The result of the fit 
is:

\begin{eqnarray}
\label{fit}
g_v^2&=&0.87883\pm0.0157\nonumber\\
\tilde{a}^{SL}&=&0.34136\pm0.0422
\end{eqnarray}    
the errors are just statistical and are obtained by increasing in one unit the 
$\chi^2$ per degree of freedom, $\chi^2_{d.o.f}$. The $\chi^2_{d.o.f}$ obtained
is:

\begin{equation}
\label{xi1}
\chi^2_{d.o.f}=0.7406
\end{equation}
with 81 experimental points.

The continuum line corresponds to the fit given by eq. (\ref{fit}). We see 
that the agreement with data is very good. Note also that $g_v^2$ is 
very close to unity. To consider the discrepancy with the KSFR result, 
which refers to the value of the coupling constant, it is better to use 
$g_v$ which results to be, from eq. (\ref{fit}), $0.94$. That is, 
only a $6\%$ of deviation respect to unity.

It is interesting and enlightening to see the value that the regularization 
scale $\mu$ should have in order to generate the $\rho$ pole at $770$ MeV when 
removing $\tilde{a}'^{L}_{\pi\pi}$ from eq. (\ref{5.2}) and setting 
$\tilde{a}^{SL}=0$. In this way we are taking the regularization scale 
as a cut off. Eq. (\ref{5.2}), neglecting $4 m_\pi^2$ with respect to 
$M_\rho^2$, transforms to:

\begin{equation}
\label{TeV}
\bigg[\frac{6 f^2}{s-4m_\pi^2}+g_0(s)  \bigg]^{-1}
\end{equation}

The resulting $\mu$ will be around by $0.7$ TeV, a value completely 
senseless \footnote{Witout neglecting
$4m_\pi^2$ compared with $M_\rho^2$ one has to multiply the quotient 
$\frac{6f^2}{s-4m_\pi^2}$ in eq. (\ref{TeV}) by
$\frac{M_\rho^2-4m_\pi^2}{M_\rho^2}\simeq 0.87$, as one can check from eq. 
(\ref{5.2}). The resulting $\mu$ is 0.3 TeV}. Its natural value is 
$\mu\simeq 1$ GeV, where typically 
resonances appear. A similar conclusion about these unrealistic high values of 
the cut-off was also obtained in \cite{Leutproce}. This 
value of $\mu$ makes it manifest that, as we have already seen in this section, 
the origin of the vectors $K^*$ and $\rho$ is attached to tree level 
structures, preexisting before unitarization. This case is different than what
will be found in the scalar case where unitarity plays a very important role.

For the scalar sector, which we will study in detail in the next section, we 
have only to make the following change in eq. (\ref{TeV}),

\begin{equation}
\label{change}
\frac{6 f^2}{s-4m_\pi^2} \longrightarrow \frac{f^2}{s-m_\pi^2/2}
\end{equation} 
that is, basically a factor 6 of difference for $s$ around $M_\rho^2$. This 
makes that the `cut-off' needed 
in the $\pi\pi$ S-wave to get a resonance of the same mass than the $\rho$ is 
just $1.8$ GeV. The change has been drastic due to the logarithmic dependence of 
the regularization scale in eq. (\ref{TeV}). We will see below the consequences 
that follow from this fact for the scalar sector.

From eqs. (\ref{rho}) and (\ref{t1n2}), one can see the limitations that some
 unitarization 
methods of the $\chi PT$ amplitudes can have when handling a general situation. In
particular one can think of the Inverse Amplitude Method (IAM) \cite{IAM1}. This
method has proved to be a very powerful tool to extend the range of
applicability of the perturbative series of $\chi PT$ up to energies around 1-1.2
GeV, giving rise to the physical resonances that appear in that case as the
$\rho$, and $K^*$ in the vector channels as well as the $\sigma$, $f_0(980)$, 
$\kappa$ and $a_0(980)$ in the scalar ones \cite{JRlargo,IAM2}. This approach is 
based in an expansion of the inverse of the K matrix \cite{JRlargo}, see eq. 
(\ref{Kmatrix}) and the sentence below it. Consider that one can neglect the 
loop contributions in the K matrix, as it occurs for the $\pi\pi$ and 
I=1/2 $K\pi$ P-wave amplitudes and it is also the case in large $N_c$ QCD. 
Then, from eq. (\ref{t1n2}), if $g_v^2\neq 1$ a zero of the tree level 
amplitude will appear, which, in turn, implies a pole in its inverse. As a 
consequence, the expansion of the inverse amplitude will break at (note that 
from eq. (\ref{Kmatrix}) a zero in the K matrix implies a zero in the full 
amplitude $\hbox{T}_L$):

\begin{equation}
\label{s0}
s_2=\frac{M_\rho^2}{1-g_v^2}
\end{equation}
where the zero of the amplitude occurs.

Thus, for our final value $g_v^2\approx 0.9$, $\sqrt{s_2}\approx 2.4$ GeV, 
further away of the physical region typically studied by this method. That 
explains the success of the IAM when applied to the vector $\pi\pi$ and I=1/2 
$K\pi$ channels in the region dominated by the $\rho$ and $K^*$ resonances, 
respectively. However, in a general scenario with $g_v^2$ very different from 
1 we cannot guarantee the applicability of the IAM. This point should be 
considered when making use of the IAM in other situations beyond QCD, as the 
ESBS \cite{RD}, where the underlying theory is not known and, furthermore, 
there are no experimental data to rely upon.

\section{The scalar sector.}

In this section we want to study the S-wave I=0,1 and 1/2 amplitudes. 
For the partial wave amplitudes with $L$=0 and I=0 and 1, coupled 
channels are fundamental in order to get an appropriate description of the 
physics involved up to $\sqrt{s}\leq1.4$ GeV. This is an important difference 
with respect the former vector channels, essentially elastic in the considered
energy region. Up to $\sqrt{s}=1.4$ GeV the most important channels are:

\begin{equation}
\label{channels}
\begin{array}{ll}
I=0 & \pi\pi(1), \; K\bar{K}(2), \; \eta\eta(3)\\
I=1 & \pi\eta(1), \;  K\bar{K}(2)\\
I=1/2& K\pi(1), \; K\eta(2) 
\end{array}
\end{equation}
where the number between brackets indicates the index associated to the 
corresponding channel when using a matrix notation as the one introduced 
in {\bf{section 2}}.

For the I=0 S-wave, the $4\pi$ state becomes increasingly important at energies 
above $1.2-1.3$ GeV, so that, in this channel, we are at the limit of 
applicability of only two mesons states when $\sqrt{s}$ is close to 1.4 GeV. 
In the I=1/2 channel, the threshold of the important $K\eta'$ state is also 
close to 1.4 GeV. Thus, one can not go higher in energies in a realistic 
description of the scalar sector without including the $K\eta'$ and 
$4 \pi$ states.

Two sets of resonances appear in the former $L$=0 partial wave amplitudes 
\cite{PDG}. A first one, with a mass around 1 GeV, contains the I=0 
$f_0(400-1200)$ and $f_0(980)$ and the I=1 $a_0(980)$. A second set appears 
with a mass around $1.4$ GeV as the I=0 $f_0(1370)$ 
and the $f_0(1500)$, the I=1 $a_0(1450)$ or the I=1/2 $K^*_0(1430)$. As a 
consequence, one could be tempted to include the exchange of two scalar 
nonets, with masses around 1 and 1.4 GeV. Before discussing whether 
this is the case, let us write the symmetric $\hbox{T}^\infty_0$ matrix 
of tree level amplitudes for the different isospins. This
matrix, $\hbox{T}^\infty_0$, is determined, as
explained at the end of {\bf{section 2}}, from the lowest order 
$\chi PT$ amplitudes,
$\hbox{T}^{(2)}$, and from the exchange of scalar nonets in the s-channel as given
by ref. \cite{EP}, $\hbox{T}^R$. In the following formulas we consider 
the exchange of only one nonet. If more nonets are needed, they 
have only to be added in the same way as the first nonet is introduced 

\vspace{1cm}

{\Large{\bf I=0}}

\begin{eqnarray}
\label{i=0}
\hbox{T}^\infty_{0,11} &=& \frac{2s-m_\pi^2}{2 f^2}+\frac{3}{2}
\frac{(\alpha_1)^2}{M_1^2-s}+\frac{3}{2}\frac{(\beta(0)_1)^2}{M_8^2-s}\nonumber \\
\hbox{T}^\infty_{0,12} &=& \frac{\sqrt{3}}{4}\frac{s}{f^2}
+\sqrt{3}\frac{\alpha_1 \alpha_2}{M_1^2-s}
+\sqrt{3}\frac{\beta(0)_1 \beta(0)_2}{M_8^2-s} \nonumber \\
\hbox{T}^\infty_{0,22} &=& \frac{3}{4}\frac{s}{f^2}
+2\frac{(\alpha_2)^2}{M_1^2-s}+2\frac{(\beta(0)_2)^2}{M_8^2-s}\nonumber \\
\hbox{T}^\infty_{0,13} &=& -\frac{m_\pi^2}{\sqrt{12}f^2}-\frac{\sqrt{3}}{2}
\frac{\alpha_1 \alpha_3}{M_1^2-s}-\frac{\sqrt{3}}{2}\frac{\beta(0)_1
 \beta(0)_3}{M_8^2-s}\nonumber\\
\hbox{T}^\infty_{0,23} &=&-\frac{3s-2 m_\eta^2-2/3 m_\pi^2}{4 f f_\eta}-
\frac{\alpha_2\alpha_3}{M_1^2-s}-\frac{\beta(0)_2\beta(0)_3}{M_8^2-s}
\nonumber \\
\hbox{T}^\infty_{0,33} &=& \frac{16 m_K^2-7 m_\pi^2}{18 f^2}+\frac{1}{2}
\frac{(\alpha_3)^2}{M_1^2-s}+\frac{1}{2}\frac{(\beta(0)_3)^2}{M_8^2-s}
\end{eqnarray}  
with $M_1$ and $M_8$ the masses of the singlet and octet in the SU(3) limit, 
$m_\eta$ is the mass of the $\eta$, 547.45 MeV \cite{PDG}, $f_\eta$ is the 
decay constant of the $\eta$, set to the value $f_\eta=1.3 f_\pi$ according 
with the $\chi PT$ prediction \cite{GL85}, $f_\pi=93.3$ MeV is
the pion decay constant and $\alpha_i$ and $\beta(0)_i$ are given by:

\begin{eqnarray}
\label{alpha0}
\beta(0)_1&=&\frac{4}{\sqrt{6}f^2}[c_d\frac{s}{2}+(c_m-c_d)m_\pi^2]
\nonumber\\
\beta(0)_2&=&-\frac{\sqrt{2}}{\sqrt{3}f^2}[c_d \frac{s}{2}+(c_m-cd)m_K^2]
\nonumber\\
\beta(0)_3&=&-\frac{4}{\sqrt{6}f^2}[c_d \frac{s}{2}+\frac{4}{3}(2c_m-c_d) 
m_K^2-(5 c_m-c_d)\frac{m_\pi^2}{3}]\nonumber\\
\alpha_1&=&\frac{4}{f^2}[\tilde{c}_d\frac{s}{2}+(\tilde{c}_m-
\tilde{c}_d)m_\pi^2]\nonumber\\
\alpha_2&=&\frac{4}{f^2}[\tilde{c}_d\frac{s}{2}+(\tilde{c}_m-
\tilde{c}_d)m_K^2]\nonumber\\
\alpha_3&=&\frac{4}{f^2}[\tilde{c}_d\frac{s}{2}+(\tilde{c}_m-
\tilde{c}_d)m_\eta^2]
\end{eqnarray}
the constants $c_d$, $c_m$, $\tilde{c}_d$ and $\tilde{c}_m$ 
characterize the coupling of a given scalar nonet to the pseudoscalar pairs of 
pions, kaons and etas as given in \cite{EP}.

\vspace{1cm}
{\Large{\bf I=1}}

\begin{eqnarray}
\label{I=1}
\hbox{T}^\infty_{0,11} &=& \frac{m_\pi^2}{3 f^2}+\frac{(\beta(1)_1)^2}
{M_8^2-s}\nonumber \\
\hbox{T}^\infty_{0,12} &=& -\frac{\sqrt{3/2}}{12 f^2}(6s-8m_K^2)
-\sqrt{2}\frac{\beta(1)_1\beta(1)_2}{M_8^2-s}\nonumber\\
\hbox{T}^\infty_{0,22} &=& \frac{s}{4 f^2}+2 \frac{(\beta(1)_2)^2}{M_8^2-s}
\end{eqnarray}
with the function $\beta(1)_i$ given by:

\begin{eqnarray}
\label{alpha1}
\beta(1)_1&=& \frac{\sqrt{2}}{\sqrt{3}f^2}[c_d(s-m_\pi^2-m_\eta^2)+
2 c_m m_\pi^2]\nonumber\\
\beta(1)_2&=& \frac{\sqrt{2}}{f^2}[c_d\frac{s}{2}+(c_m-c_d)m_K^2]
\end{eqnarray}

\vspace{1cm}
{\Large{\bf I=1/2}}

\begin{eqnarray}
\label{I=1/2}
\hbox{T}^\infty_{0,11} &=&\frac{5s^2-2s(m_\pi^2+m_K^2)-3(m_K^2-m_\pi^2)^2}
{8s f_K^2}+
\frac{3}{2}\frac{(\beta(1/2)_1)^2}{M_8^2-s}\nonumber\\
\hbox{T}^\infty_{0,12} &=&\frac{-9s^2+2sm_K^2+3sm_\eta^2+7sm_\pi^2-9m_K^4
+9m_K^2(m_\pi^2+m_\eta^2)-9m_\pi^2 m_\eta^2}{24 s f_K^2}+\sqrt{\frac{3}{2}}
\frac{\beta(1/2)_1\beta(1/2)_2}{M_8^2-s}\nonumber\\
\hbox{T}^\infty_{0,22}&=&\frac{-9s^2-9(m_k^2-m_\eta^2)^2+6s(3m_K^2+m_\eta^2)-
4sm_\pi^2}{24 s f_K^2}+\frac{(\beta(1/2)_2)^2}{M_8^2-s}
\end{eqnarray}
with $f_K$ the kaon decay constant with the value $f_K=1.2f_\pi$ according 
to experiment \cite{36}. The functions $\beta(1/2)_i$ are given 
by

\begin{eqnarray}
\label{alpha12}
\beta(1/2)_1&=&\frac{1}{f_K^2}[c_d s + (c_m-c_d)(m_K^2+m_\pi^2)]\nonumber\\
\beta(1/2)_2&=&-\frac{1}{\sqrt{6}f_K^2}[c_ds+c_m(5m_K^2-3m_\pi^2)
-c_d(m_K^2+m_\eta^2)]
\end{eqnarray}

Note that the introduction of a nonet implies six new parameters, two masses 
and four coupling constants, which we fit to the experiment.

According to eq. (\ref{TL}), we also need the function $g_0(s)$, given by 
eqs. (\ref{g(s)dr}) and (\ref{g(s)dr2}), with its corresponding $a^{SL}$ for the S-wave channels. 
By SU(3) arguments, the $a^{SL}$ constant can be different for vector and 
scalar channels. The reason is that a two meson state has different SU(3) wave 
functions in S and P wave, because under the exchange of both 
mesons the spatial P-wave is antisymmetric while the S-wave is symmetric and the
total wave function must be symmetric. 
That is, the two mesons are in different SU(3) representations.

We have included $f_\eta$ in the S-wave isoscalar ${\mathcal{O}}(p^2)$ 
$\chi PT$ amplitude for $K\bar{K}\rightarrow \eta\eta$ and $f_K$ in the 
S-wave, I=1/2 tree level amplitudes, to obtain, after the fit, that the $a^{SL}$ 
constant was the same for all the scalar channels. These changes come from the 
SU(3) breaking of the octet of ($\pi$,$K$,$\eta$) and can not be taken into 
account, a priori, in our way of fixing $D_L(s)$ making use of lowest order 
$\chi PT$ \cite{Retaila} and the exchange of resonances given by \cite{EP}.

The fit will be done for the following experimental data: the elastic S-wave 
$\pi\pi$ phase shifts with 
I=0, $\delta^{00}_{11}$, the $K\bar{K}\rightarrow \pi\pi$, I=$L$=0 phase 
shifts, $\delta^{00}_{12}$, the I=$L$=0 $\frac{1-\eta^2_{00}}{4}$, with
$\eta_{00}$ the inelasticity in that channel, the elastic S-wave, I=1/2 
$K\pi$ phase shifts, $\delta^{\frac{1}{2}\,0}_{11}$ and a distribution of 
events around 
the mass of the $a_0(980)$ resonance, corresponding to the central production 
of $\pi\pi\eta$ in 300 GeV $pp$ collisions \cite{Arm} for the I=1, $L$=0 
channel.

\vspace{0.5cm}
{\bf{$\delta^{00}_{11}$}}
\vspace{0.2cm}

Because results coming from different experiments analyses are not 
compatible, we have taken as central value for each energy below $\sqrt{s}
= 1$ GeV the mean between the different experimental results \cite{38,39}. 
For $\sqrt{s}>1$ GeV, the mean value comes from \cite{39,40}. In both cases 
the error is the maximum between the experimental one and the largest 
distance of the experimental values to the mean one.  This procedure 
will be the one adopted, when needed, for the rest of the experimental 
magnitudes included in the fit.

\vspace{0.5cm}
{\bf{$\delta^{00}_{12}$}}
\vspace{0.2cm}

For this quantity there are two sets of data below $\sqrt{s}=1.2$ GeV. Higher 
in energy both sets converge. One group will be represented by \cite{42} and 
the other one by \cite{43,44}. The experimental results from \cite{42} are 
larger than the data of the other works \cite{43,44} below 1.2 GeV. We will 
distinguish between both cases when doing a fit referring it as 
{\bf{high/below}} respectively. The change in the value of the fitted 
parameters will 
be very small when changing from one set of data to another, so that, this 
experimental ambiguity will not be relevant for our final values. We will 
average the experimental data of the second set of works for 
$\sqrt{s}\leq1.2$ GeV in the way explained above. When $\sqrt{s}>1.2$ GeV the 
average will be done between all the quoted analyses \cite{42,43,44}.

\vspace{0.5cm}
{\bf{$\frac{1-\eta^2_{00}}{4}$}}
\vspace{0.2cm}

There are a series of analyses and experiments about the 
inelastic cross section $\pi\pi \rightarrow K\bar{K}$, which agree between 
each other in the values for $\frac{1-\eta^2_{00}}{4}$. We have taken the data
from \cite{42,43} as representative for such situation. 

The quantity $\frac{1-\eta^2_{00}}{4}$ has been used instead of the
inelasticity, $\eta_{00}$, because the former is much better measured and 
all the experiments \cite{42,43,44} agree on that quantity.

\vspace{0.5cm}
{\bf{$\delta^{\frac{1}{2} \,0}_{11}$}}
\vspace{0.2cm}

We distinguish between the more recent experiment \cite{45} and the older
results \cite{36o,390,47}. We have averaged the data from the last analyses 
up to $\sqrt{s}=1$ GeV. Above this energy, in the latter group of experimental 
works, only \cite{390} offers data. The statistical errors in this latter
experiment are very small. We have enlarged them at the level of those in the
most recent experiment \cite{45} which would make the different experiments
compatible.
Thus, the final points used in the fit for this magnitude will be the ones 
from \cite{45} and the average between \cite{36o,390,47} as described above.

\vspace{0.5cm}
I=1, $L$=0 data.
\vspace{0.2cm}

The experimental data is very scarce for this channel. We will take a
distribution of events corresponding to the central production of $\pi\pi\eta$
in 300 GeV $pp$ collisions \cite{Arm}. We will study the data points around the
mass of the $a_0(980)$, where one could think that the energy dependence will
be dominated by the exchange of that resonance. We add in an incoherent way with
respect the $a_0(980)$ resonance, the same background as in \cite{Arm}. The 
$a_0(980)$ contribution is parametrized as:

\begin{equation}
\label{a0dis}
\frac{\hbox{d}\hbox{N}}{\hbox{d}E_{cm}}={\mathcal{N}}\, p_{\pi\eta}\, 
|(\hbox{T}_{0})_{12}|^2
\end{equation} 
with $p_{\pi\eta}$ the three momentum of the $\pi\eta$ state in the CM
corresponding to a total energy $E_{cm}$  and 
${\mathcal{N}}$ is just a normalization constant.

\vspace{1cm}
{\bf{The fit.}}
\vspace{0.4cm}

Let us now discuss the fits obtained when including: (1) two scalar nonets with
masses around 1 and 1.4 GeV or (2) only one nonet with a mass around 1.4 GeV. 
Of course, the final value for the tree level, `bare', masses of the octet and 
singlet will be given by the fit. The fits have been done using the MINUIT
minimization program. The output value for the parameters are written with the
same precision as given by MINUIT.

The fit that results when two nonets are included and also with the high
$\delta^{00}_{12}$ data is:

\begin{equation}
\label{fit1}
\begin{array}{lll}
\hbox{First Nonet (MeV)} & \hbox{Second Nonet (MeV)} & \\
c_d=1.7997 & c'_d=19.512 & a^{SL}=-.72225\\
c_m=0.66534& c'_m=19.612& {\mathcal{N}}=9.2195\;\, \hbox{MeV}^{-2}\\
M_8=1003.41\pm 600 & M'_8=1379.31 &\\
\tilde{c}_d=20.988 & \tilde{c}'_d=0.32994&\\
\tilde{c}_m=8.4867 & \tilde{c}'_m=-2.7183&\\
M_1=1031.91 & M'_1=1000 \pm 600&\\
&&\\
\chi^2_{d.o.f}=0.9688&&\\
188 \;\, \hbox{points}
\end{array}
\end{equation}

A very striking aspect appears when observing the value of the parameters 
given in eq. (\ref{fit1}). The
value of the constants $c_d$, $c_m$ and $\tilde{c}'_d$ and $\tilde{c}'_m$ are,
at least, one order of magnitude smaller than $\tilde{c}_d$, $\tilde{c}_m$ and 
$c'_d$, $c'_m$ respectively. This makes that the first octet and second
singlet are phenomenologically irrelevant. Note that their masses are
essentially undetermined. This is shown by the need to increase them by 600 MeV 
in order to make the $\chi^2_{d.o.f}$ to increase by 0.5 units. In this way, they 
do not originate or participate in the poles corresponding to the physical
 resonances mentioned at the beginning of the section. They only give rise to 
 poles very close to the real axis, with a width of only a few MeV. These 
poles manifest themselves as very narrow peaks in the partial waves, which
are not observed by experiment.

From the latter discussion we will just introduce a scalar nonet. The resulting 
values, after a new fit to the data, are:

\vspace{0.3cm}
{\bf{High $\delta^{00}_{12}$}}

\begin{equation}
\label{fit2}
\begin{array}{ll}
\hbox{Nonet (MeV)} & \\
c_d=19.113^{+2.4}_{-2.1} & a^{SL}=-.75110\pm 0.2\\
c_m=15.110\pm 30& {\mathcal{N}}=9.3922 \pm 4.5\;\,\hbox{MeV}^{-2}\\
M_8=1390.31\pm 20  &\\
\tilde{c}_d=20.918^{+1.6}_{-1.0} &   \chi^2_{d.o.f}=1.066 \\
\tilde{c}_m=10.567^{+4.5}_{-3.5} & 188 \;\, \hbox{points}\\
M_1=1021.11^{+40}_{-20}&
\end{array}
\end{equation}

\vspace{0.3cm}
{\bf{Low $\delta^{00}_{12}$}}

\begin{equation}
\label{fit3}
\begin{array}{ll}
\hbox{Nonet (MeV)} & \\
c_d=19.183  & a^{SL}=-.74246 \\
c_m=15.248 & {\mathcal{N}}=9.429\;\,\hbox{MeV}^{-2}\\
M_8=1390 & \\
\tilde{c}_d=20.941 & \chi^2_{d.o.f}=1.21 \\
\tilde{c}_m=10.641 &196 \;\, \hbox{points} \\
M_1=1021 &
\end{array}
\end{equation}

We have also shown the statistical errors
for the parameters of the high $\delta^{00}_{12}$ fit
obtained by increasing the $\chi^2_{d.o.f}$ by one unit, in order to appreciate
the precision in the value of the parameters given by the last fits. The large
error on $c_m$ is because this constant, as can be seen from eqs. 
(\ref{alpha0}), (\ref{alpha1}) and (\ref{alpha12}), enters through the 
multiplication of
squared masses of the lightest pseudoscalars which are much smaller than
$s\simeq M_8^2$, around the resonance region of the octet. Thus, its influence
in the final value of the amplitudes is very small. This also happens to 
$\tilde{c}_m$, although to a lower extension because $M_1<M_8$. One can 
see, comparing the
last two fits, that the variation in the value of the parameters is very small
when changing from one set of data to the other. The resulting fit for the high
$\delta^{00}_{12}$ data is shown in Figs. 4-8. The results obtained before also
favor the high solution for the $\delta_{12}^{00}$ phase shifts because its
corresponding $\chi^2_{d.o.f}$ is smaller than the one for the low
$\delta^{00}_{12}$ solution.

\begin{figure}[h]
\centerline{
\protect
\hbox{
\psfig{file=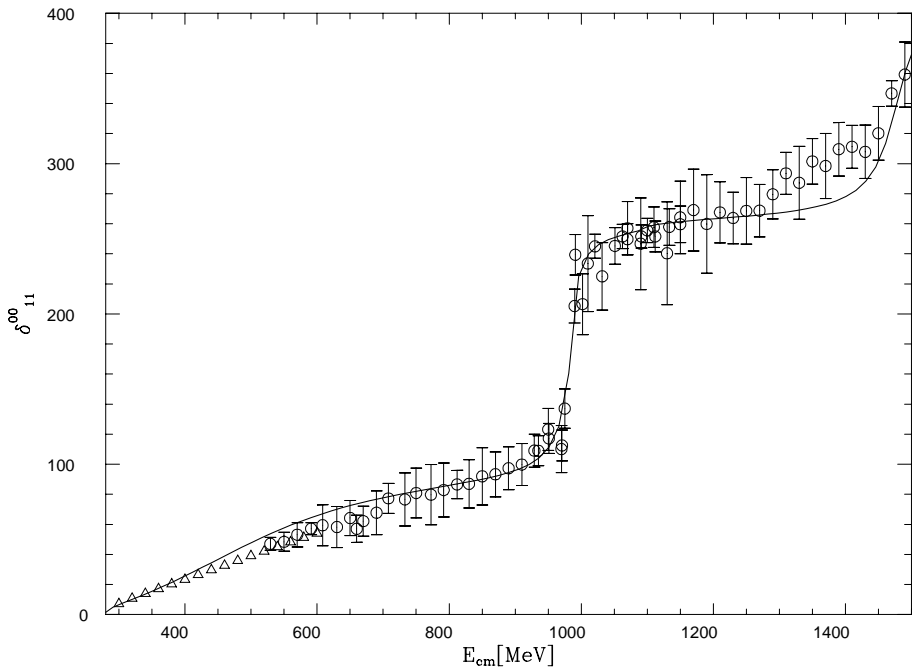,width=0.6\textwidth,angle=-90}}}
\caption{Elastic isoscalar $\pi\pi$ phase shifts, $\delta^{00}_{11}$. The
circles correspond to the average of \cite{38,39,40}, as discussed in the
$\delta^{00}_{11}$ subsection. We have also included the triangle points form
\cite{41} to have some data close to threshold, although these
points have not been included in the fit because they are given without errors.}
\end{figure}   

\begin{figure}[h]
\centerline{
\protect
\hbox{
\psfig{file=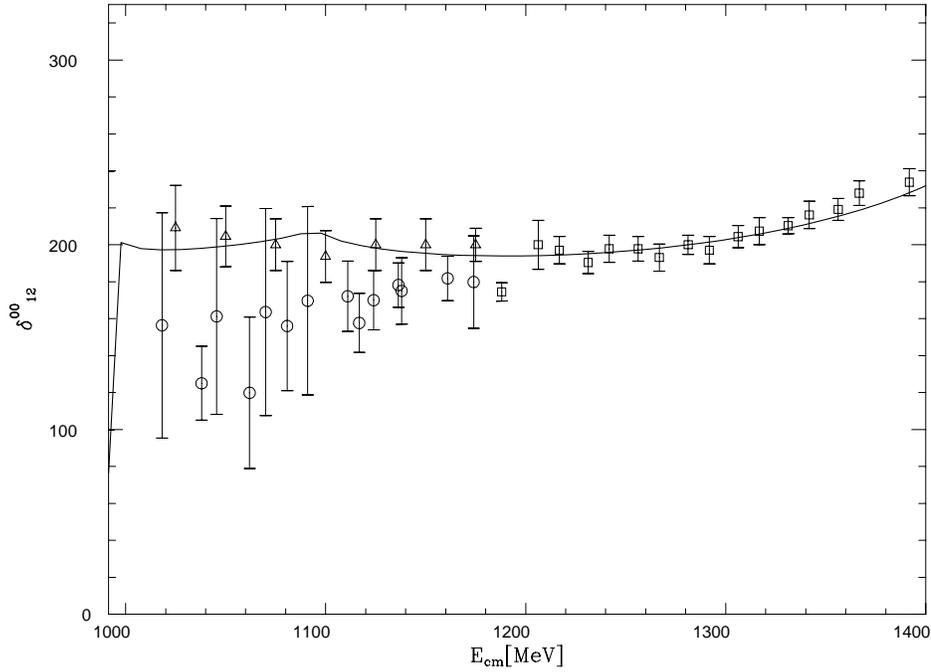,width=0.6\textwidth,angle=-90}}}
\caption{S-wave $K\bar{K}\rightarrow \pi\pi$ isoscalar phase shifts,
$\delta^{00}_{12}$. The triangles points are from \cite{42}, circles correspond
to the average of \cite{43,44} and squares to the one of \cite{42,43,44}.}
\end{figure}

This fit has 8 free parameters, 6 constants from the nonet\footnote{If the 
singlet and the octet introduced form really a nonet is something we can not
say. However, we will denote the global contribution of the introduced 
octet plus the singlet by using the word nonet as a shortcome. In this way, we 
 also follow the nomenclature of \cite{EP}, inspired in the U(3) symmetry 
 which holds for $N_c \rightarrow \infty$ .}, $a^{SL}$ and the normalization 
constant
${\mathcal{N}}$. In our former work \cite{48}, we were able to describe
$\delta^{00}_{11}$, $\delta^{00}_{12}$ and $\frac{1-\eta^2_{00}}{4}$ up to
$\sqrt{s}\leq1.2$ GeV and a
distribution of events around the $a_0(980)$ mass \cite{Flate}, using only 2
free parameters: a cut-off (which plays the role of the regularization
scale $\mu$ and at the same time generates a concrete value for $a^{SL}$) and a
corresponding normalization constant for the $a_0(980)$ event distribution. In
fact, if we remove the resonance contributions in eqs. (\ref{i=0}) and 
(\ref{I=1}) the formalism of ref. \cite{48} follows from the present one. It
might look surprising that a good fit to the data for $\sqrt{s}<1.2$ GeV could
be obtained in \cite{48} with just one free parameter and a normalization
constant for the mass distribution, while here one has needed 7
parameters, apart from the normalization constant. One reason is that now 
we have pushed the fit up to $\sqrt{s}=1.4$
GeV, while in \cite{48} only data up $\sqrt{s}=1.2$ GeV was considered. The fact
that new resonances appear around $\sqrt{s}=1.4$ GeV has forced us to include an
octet, which implies 3 new parameters, two couplings and a mass. However, the
effect of this octet below $\sqrt{s}=1.2$ GeV is very small, hence only the
singlet appearing with a mass around 1 GeV is relevant for energies below
$\sqrt{s}=1.2$ GeV. The present fit to the data has led us to the inclusion of
this singlet resonance in $\hbox{T}^\infty_0$ apart from the lowest order $\chi
PT$ Lagrangian, while in \cite{48} only the latter contribution was considered.
The reason that forced us to include now the mentioned singlet is the
consideration of the $\eta\eta$ channel in I=0, which was omitted in \cite{48},
and is not negligible above 1 GeV, as can be seen in its strong coupling to the
$f_0(980)$ resonance that we will obtain below. The $\eta\eta$ 
channel affects mostly the magnitude $\frac{1-\eta_{00}^2}{4}$. Should one have
taken the available data for $\eta_{00}$ instead of those for 
$\frac{1-\eta_{00}^2}{4}$, which are measured with better precision from the
$\pi\pi \rightarrow K\bar{K}$ inelastic cross section, the effect of the
$\eta\eta$ channel would be masked by the large errors in $\eta_{00}$.

It is quite interesting to recall that an I=0 elementary state around 1 GeV has
been predicted from QCD inspired models \cite{Peris,Espriu} and 
has also been advocated in phenomenological
analyses \cite{Au,Penotros}. Such state could be associated with the preexisting
singlet state that we need.

\vspace{0.5cm}
{\bf{Resonances.}}
  
Let us now concentrate on the resonance content of the fit presented in eq. 
(\ref{fit2}). The octet around 1.4 GeV gives rise to eight resonances which
appear with masses very close to the physical ones, $f_0(1500)$, $a_0(1450)$ 
and $K^*_0(1430)$ \cite{PDG}. Thus, the correlation between tree level 
resonances, poles and physical resonances is clear around $\sqrt{s}\simeq 1.4$ 
GeV. However, this correlation is not so clear around 1 GeV. This issue will be 
the object of the following discussions \footnote{We do not give a detailed
study for the resonances with masses around 1.4 GeV because we have not included
channels which become increasingly important for energies above $\simeq$ 1.3 
GeV as $4\pi$ in I=0 or $K\eta'$ for I=1/2. This makes that the widths we obtain
from the pole position of the former resonances are systematically smaller than
the experimental ones \cite{PDG}. Thus, a more detailed study, which
included all the relevant channels for energies above 1.3 GeV, should be done 
in order to obtain a better determination of the parameters for this octet 
around 1.4 GeV.}.

From Figs. 4 and 7, one can easily see two resonances with masses around 1
GeV, the well known $f_0(980)$ and $a_0(980)$ resonances. The first one could
be related to the singlet bare state with $M_1=1020$ MeV, but for the second we
have not bare resonances to associate with, because the tree level resonance
was included with a mass around 1.4 GeV and has evolved to the physical
$a_0(1450)$. The situation is even more complex, because we also find in our 
amplitudes other
poles corresponding to the $f_0(400-1200)\equiv \sigma$ and to the
$K^*_0(900)\equiv \kappa$. In Table 1 the pole positions of the resonances in
the second sheet\footnote{I sheet: Im $p_1>$0, Im $p_2>$0, Im $p_3>$0; II 
sheet: Im $p_1<$0, Im $p_2>$0, Im $p_3>$0} are given and also the modulus of 
the residues
corresponding to the resonance $R$ and channel $i$, $\zeta^R_i$, given by

\begin{equation}
\label{residue}
|\zeta^R_i \zeta^R_j|=\lim_{s \to s_R} |(s-s_R) \hbox{T}_{ij}|
\end{equation}
where $s_R$ is the complex pole for the resonance $R$.

\begin{table}
\begin{center}
\caption{Pole position and residues for the full amplitude.}
\begin{tabular}{|c|c|}     
\hline
&\\
\begin{math}
\begin{array}{ccc}
\sqrt{s}_{\sigma}&=&445+i\; 221 \;\, \hbox{MeV}\\
&&\\
\zeta^{\sigma}_{\pi\pi}&=&4.26 \;\, \hbox{GeV}\\
&&\\
\frac{\zeta^{\sigma}_{K\bar{K}}}{\zeta^{\sigma}_{\pi\pi}}&=&0.254\\
&&\\
\frac{\zeta^{\sigma}_{\eta\eta}}{\gamma^{\sigma}_{\pi\pi}}&=&0.036 
\end{array}
\end{math} &
\begin{math}
\begin{array}{ccc}
\sqrt{s}_{f_0}&=&987+i\; 14 \;\, \hbox{MeV}\\
&&\\
\zeta_{K\bar{K}}^{f_0}&=&3.63\;\, \hbox{GeV}\\
&&\\
\frac{\zeta_{\pi\pi}^{f_0}}{\zeta_{K\bar{K}}^{f_0}}&=&0.51\\
&&\\
\frac{\zeta_{\eta\eta}^{f_0}}{\zeta_{K\bar{K}}^{f_0}}&=&1.11
\end{array}
\end{math}\\
&\\
\hline
&\\
\begin{math}
\begin{array}{ccc}
\sqrt{s}_{{a_0}}&=&1053.13+i\; 24 \;\, \hbox{MeV}\\
&&\\
\zeta_{K\bar{K}}^{{a_0}}&=&5.48 \;\, \hbox{GeV}\\
&&\\
\frac{\zeta_{\pi\eta}^{{a_0}}}{\zeta_{K\bar{K}}^{{a_0}}}&=&0.70
\end{array}
\end{math}&
\begin{math}
\begin{array}{ccc}
\sqrt{s}_{\kappa}&=&779+i\; 330 \;\, \hbox{MeV}\\
&&\\
\zeta_{K\pi}^{\kappa}&=&4.99 \;\, \hbox{GeV}\\
&&\\
\frac{\zeta_{K\eta}^{\kappa}}{\zeta_{K\pi}^{\kappa}}&=&0.62
\end{array}
\end{math}\\
&\\
\hline
\end{tabular}

\caption{Pole position and residues when the bare resonant contributions are
removed}
\begin{tabular}{|c|c|}     
\hline
&\\
\begin{math}
\begin{array}{ccc}
\sqrt{s}_{\sigma}&=&434+i\; 244 \;\, \hbox{MeV}\\
&&\\
\zeta^{\sigma}_{\pi\pi}&=&4.21 \;\, \hbox{GeV}\\
&&\\
\frac{\zeta^{\sigma}_{K\bar{K}}}{\zeta^{\sigma}_{\pi\pi}}&=&0.301\\
&&\\
\frac{\zeta^{\sigma}_{\eta\eta}}{\gamma^{\sigma}_{\pi\pi}}&=&0.033 
\end{array}
\end{math} &
\begin{math}
\begin{array}{ccc}
\sqrt{s}_{f_0}&=&\hbox{cusp effect}\\
&&\\
\zeta_{K\bar{K}}^{f_0}&=& ... \\
&&\\
\frac{\zeta_{\pi\pi}^{f_0}}{\zeta_{K\bar{K}}^{f_0}}&=&0.38\\
&&\\
\frac{\zeta_{\eta\eta}^{f_0}}{\zeta_{K\bar{K}}^{f_0}}&=&1.04
\end{array}
\end{math}\\
&\\
\hline
&\\
\begin{math}
\begin{array}{ccc}
\sqrt{s}_{{a_0}}&=&1081.95+i\; 13.3 \;\, \hbox{MeV}\\
&&\\
\zeta_{K\bar{K}}^{{a_0}}&=&5.98 \;\, \hbox{GeV}\\
&&\\
\frac{\zeta_{\pi\eta}^{{a_0}}}{\zeta_{K\bar{K}}^{{a_0}}}&=&0.74
\end{array}
\end{math}&
\begin{math}
\begin{array}{ccc}
\sqrt{s}_{\kappa}&=&770+i\; 341 \;\, \hbox{MeV}\\
&&\\
\zeta_{K\pi}^{\kappa}&=&4.87 \;\, \hbox{GeV}\\
&&\\
\frac{\zeta_{K\eta}^{\kappa}}{\zeta_{K\pi}^{\kappa}}&=&0.61
\end{array}
\end{math}\\
&\\
\hline
\end{tabular}
\end{center}
\end{table}

\begin{figure}[h]
\centerline{
\protect
\hbox{
\psfig{file=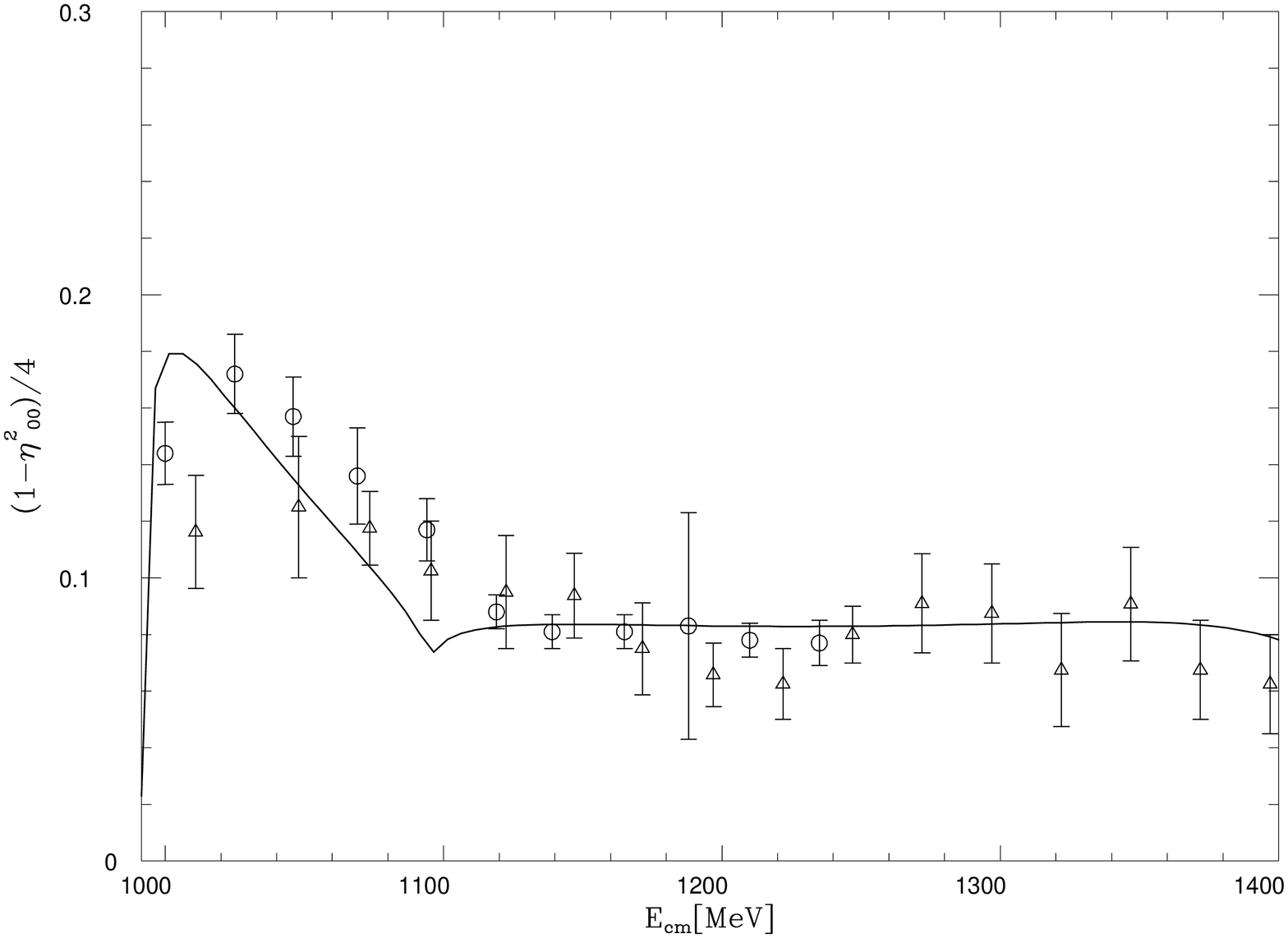,width=0.6\textwidth,angle=-90}}}
\caption{$\frac{1-\eta^2_{00}}{4}$ with $\eta_{00}$ the I=$L$=0 S-wave
inelasticity. Circles \cite{42}, triangles \cite{43}.}
\end{figure}

While for the $f_0(980)$ one has a preexisting tree level resonance with a mass
of 1020 MeV, for the other resonances present in Table 1 the situation is rather
different. In fact, if we remove the tree level nonet contribution from eqs.
(\ref{i=0}), (\ref{I=1}) and (\ref{I=1/2}) the $a_0(980)$, $\sigma$ and
$\kappa$ poles still appear as can be seen in Table 2. For the $f_0(980)$, in
such a situation, one has not a pole but a very strong cusp effect in the
opening of the $K\bar{K}$ threshold. In fact, by varying a little the
value of $a^{SL}$ one can regenerate also a pole for the $f_0(980)$ from this
strong cusp effect. In Table 2 we have not given an absolute value for the
coupling of the $f_0(980)$ to the $K\bar{K}$ channel because one has not a pole
for the given value of $a^{SL}$. However, the ratios between the different amplitudes
are stable around the cusp position. As a result, the physical $f_0(980)$ will 
have two contributions: one from the bare singlet state with $M_1=1020$
MeV and the other one coming from meson-meson scattering, particularly
$K\bar{K}$ scattering, generated by the lowest order $\chi PT$ Lagrangian.

In eqs. (\ref{i=0}), (\ref{I=1}) and (\ref{I=1/2}) when the resonant
tree level contributions are removed, only the lowest order, 
${\mathcal{O}}(p^2)$, $\chi PT$ contributions remain. Thus, except for the
contribution to the $f_0(980)$ coming from the bare singlet at 1 GeV, the poles
present in Table 2 originate from a `pure potential' scattering, following the
nomenclature given in \cite{Chew1}. In this way, the source of the dynamics is 
the lowest order $\chi PT$ amplitudes. The constant $a^{SL}$ can be
interpreted from the need to give a `range' to this potential so that the loop 
integrals converge. These meson-meson states are shown in Fig. 9 in the
chiral limit, setting all the masses of the pseudoscalars to zero and all the
$f_P=f$, where $P$ denotes any pseudoscalar meson $\pi$, $K$ or $\eta$. We see 
in this last figure a degenerate octet for I=0,1 and 1/2
with a mass around 500 MeV and a singlet in I=0 with 400 MeV of mass. In both
cases these meson-meson resonances are very broad.

The situation is very different to that of the former studied vector channels 
where all the physical resonances, the $\rho$ and $K^*$, originate from the 
preexisting tree level resonances. We already saw, at the end of the last 
section, when comparing the S and P-wave $\pi\pi$ scattering, that the 
${\mathcal{O}}(p^2)$ $\chi PT$ amplitude is 6 times 
larger for $L$=0 than for $L$=1 around the resonance energy region. This implies 
that n-loops in $L$=1, with the 
${\mathcal{O}}(p^2)$ $\chi PT$ amplitudes at the vertices, will be suppressed 
by a factor ${\displaystyle{\frac{1}{6^{n+1}}}}$ with 
respect to $L$=0. The suppression of loops is expected from Large 
$N_c$ QCD and this is in fact what happens for the vector channels, but for 
the scalar ones unitarity is unexpectly large, giving rise to these meson-meson 
resonances.  

As can be seen from eq. (\ref{TL}), these meson-meson poles, without tree level
resonant contributions, originate from
the cancellation between the inverse of the ${\mathcal{O}}(p^2)$ $\chi PT$ 
amplitude and the $g_0$ function. As a consequence, the following relation
between the masses of those resonances with $f$ results

\begin{equation}
\label{MLN}
M^2 \propto f^2/g_0
\end{equation}
since $g_0$ is ${\mathcal{O}}(1)$ and $f^2$ is ${\mathcal{O}}(N_c)$
\cite{Witten}, these masses will grow as $N_c$. Thus for $N_c\rightarrow\infty$
these resonances will go to infinity. This movement can be followed by
suppressing the $g_0$ function by a factor $\tau$ from 1 (physical situation)
to 0 ($N_c=\infty$). It is then observed how the resonances in Table 2, without
the preexisting resonant contributions, disappear going to infinity.

\begin{figure}[h]
\centerline{
\protect
\hbox{
\psfig{file=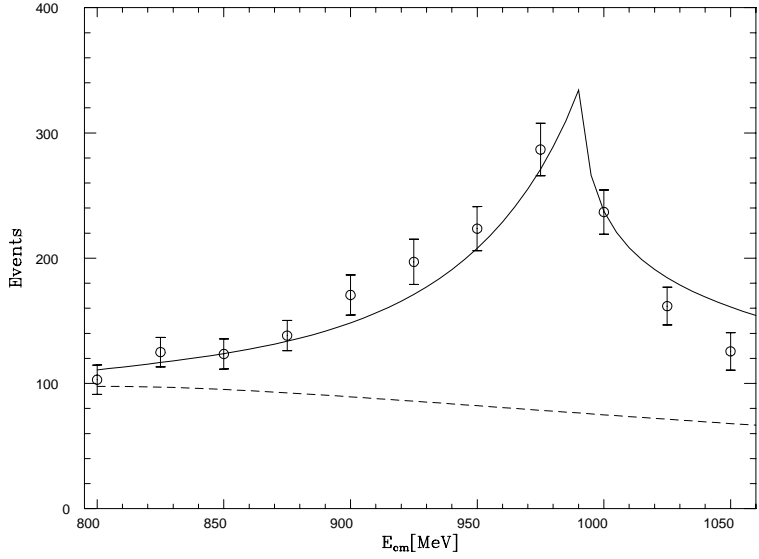,width=0.5\textwidth,angle=-90}}}
\caption{Distribution of events around the $a_0(980)$ mass corresponding to the
central production $\pi\pi\eta$ in 300 GeV collisions \cite{Arm}. The abscissa 
represents the $\pi \eta$ invariant mass, $E_{cm}$. The dashed
line represents the background introduced in the same reference.}
\end{figure}

\begin{figure}[H]
\centerline{
\protect
\hbox{
\psfig{file=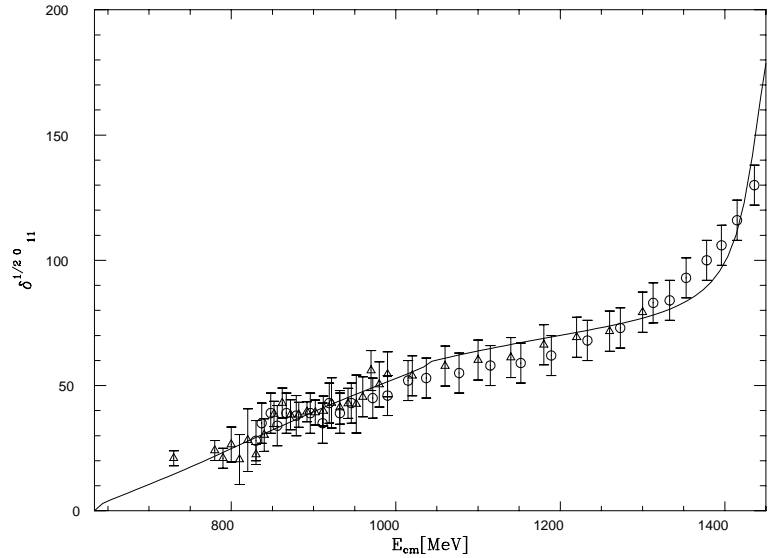,width=0.5\textwidth,angle=-90}}}
\caption{S-wave I=1/2 $K\pi$ elastic phase shifts, $\delta^{1/2 \,0}_{11}$. The
triangles correspond to the average, as described in $\delta^{1/2 \,0}_{11}$
subsection, of \cite{36o,390,47}. Circles correspond to \cite{45}.}
\end{figure}

\begin{figure}[H]
\hbox{
\epsfig{file=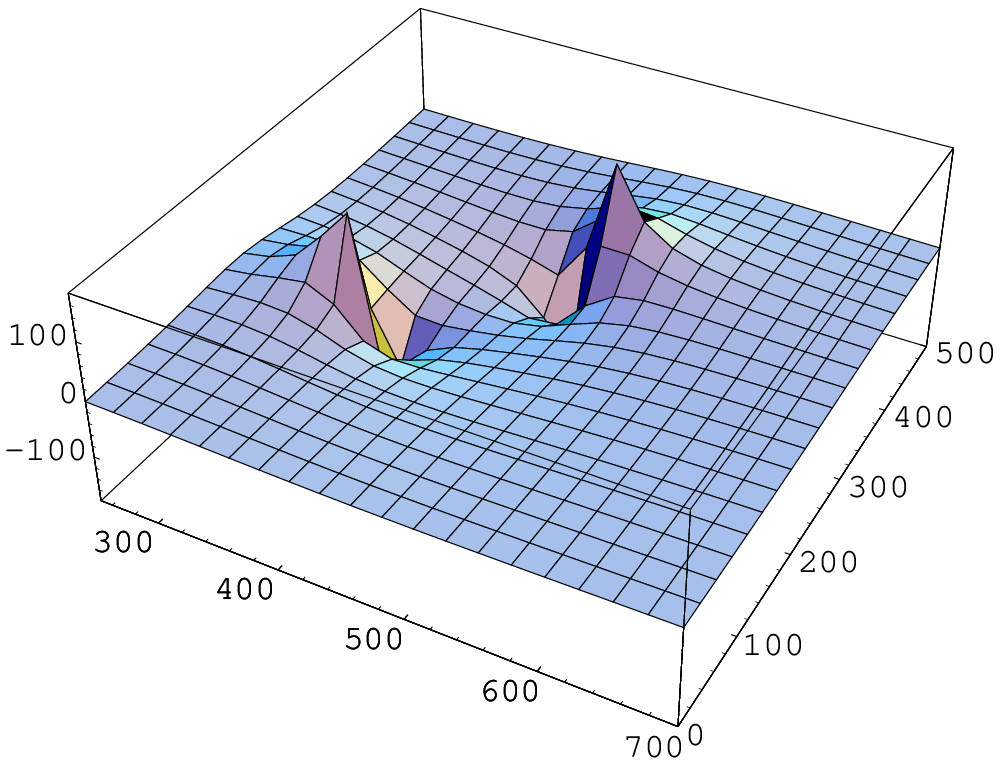,width=5cm}
\epsfig{file=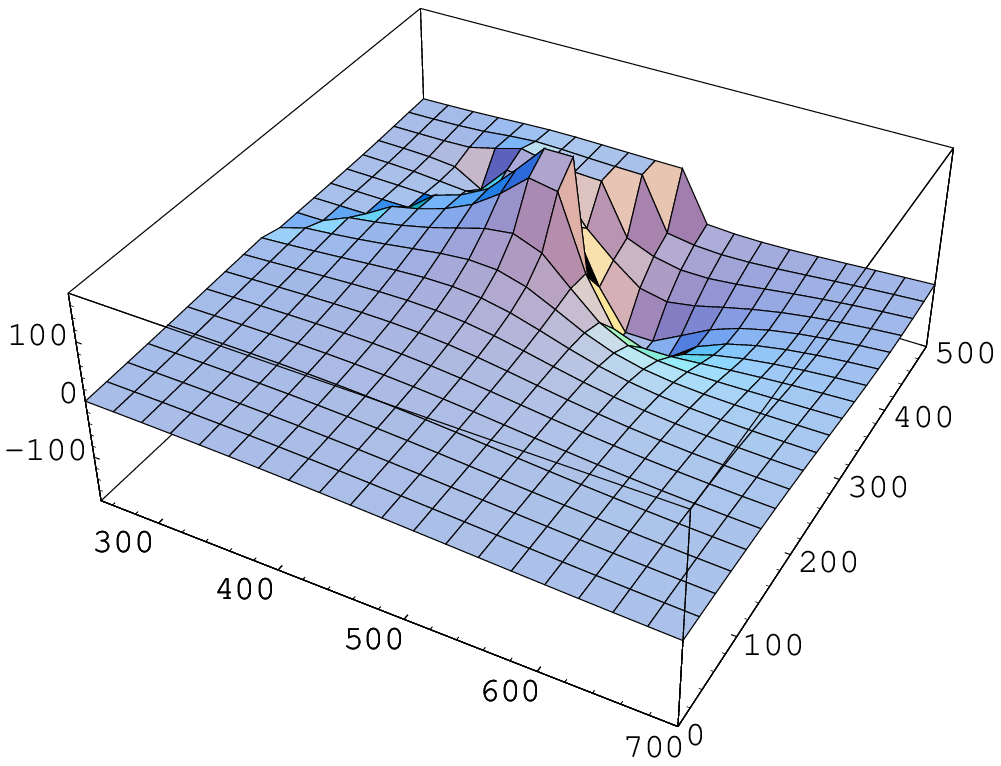,width=5cm}
\epsfig{file=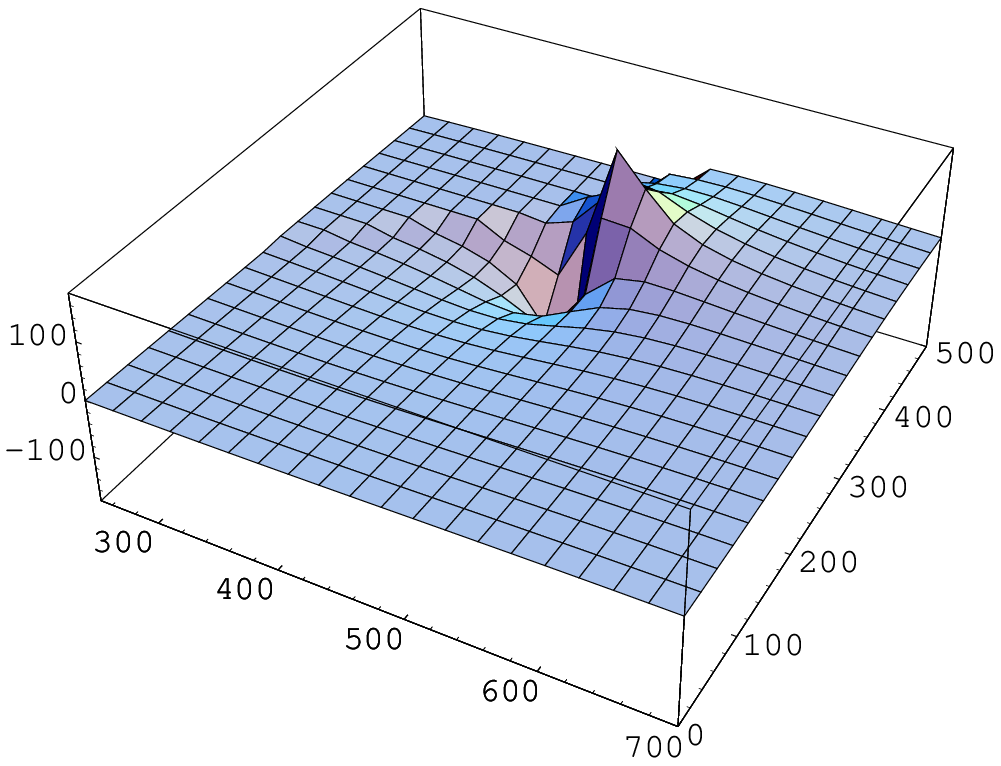,width=5cm}}
\caption{{\bf{Chiral limit}}. From left to right, Figs. 9a,b,c 
respectively. In Fig. 9a the poles of T found in the unphysical sheet are 
shown for I=0. Analogously for I=1/2 and 1 in Figs. 9b,c respectively.}  
\end{figure}  

\section{Estimations of the unphysical cut contribution from $\chi$PT and the 
exchange of resonances.}	

	In this last section we estimate the influence of the 
unphysical cuts for the elastic $\pi\pi$ and $K\pi$ S-waves with I=0 and 1/2
respectively. The unphysical cuts will be approximated by means of $\chi$PT 
supplied with the exchange of resonances \cite{EP} with spin $\leq$1 in the 
t- and u-channels. The loops are calculated from 
$\chi$PT at $\mathcal{O}(p^4)$ and the exchange of resonances in the crossed 
channels accounts for a 
resummation of counterterms up to an infinite order, in the way explained 
in {\bf{section 2}} after eq. (\ref{limitd'}). The result can be taken directly 
from ref. \cite{UMR}, where the $\pi\pi\rightarrow \pi\pi$ and $K\pi
\rightarrow K\pi$ amplitudes are calculated up to one loop including 
explicit resonance fields \cite{EP}. 

	In order to extract, from ref. \cite{UMR}, the contribution of the 
unphysical cuts, which we design by $\hbox{T}_{Left}$, we have made use of eqs. 
(3.2), (3.10), (3.13), (3.14) for the $\pi\pi$ scattering and of eqs. (3.6), 
(3.16), (3.19) and (3.20) for the $K\pi$ one. We calculate the loop
contributions at the same regularization scale than in \cite{UMR}, that is,
$\mu=M_\rho=770$ MeV, the same we have taken in this work. In the following, when we 
refer to an equation in the form (m.n), it should 
be understood that this equation is the corresponding one from ref. \cite{UMR}. 
The work of \cite{UMR} contains loops and the exchange of resonances in the $s$,
$t$ and $u$ channels. The exchange of resonances in the s-channel is also
present in our work where the masses and couplings of the scalar resonances,
eqs. (\ref{fit2}), were fitted to data. Obviously the loops and the exchange of 
resonances in crossed channels, absent in our work, go to $\hbox{T}_{Left}$. On the
other hand, we must also include in $\hbox{T}_{Left}$ a polynomial contribution
of $\mathcal{O}(p^4)$ because the loop functions used in \cite{UMR}, $J_{PQ}^r$,
 $M_{PQ}^r$ and $\bar{J}_{PQ}$, where $P$, $Q$ are $\pi$, $K$ or $\eta$, and our
 loop functions, $(-)g_0(s)$, differ in a constant. This polynomial contributions
 can be interpreted as subtraction terms from a dispersion relation of 
 $\hbox{T}_{Left}$.  Since the loops are calculated with $\hbox{ImT}$ at
 $\mathcal{O}(p^4)$ one needs three subtractions, which fix the order of the
 subtraction polynomial. Let us explain first the $\pi\pi$ scattering. 
  
	From eq. (3.2) one has the expression of the elastic 
$\pi\pi$ amplitude with I=0 in terms of the amplitude A$(s,t,u)$, eq. (3.10).
 Making use of eqs. (3.2) and (3.13) the 
contribution of the loops in the s-channel is given by

\begin{equation}
\label{pipi1}
\frac{(2s-m_\pi^2)^2}{2 f_\pi^4} J^r_{\pi\pi}(s)+\frac{3 s^2}{8 f_\pi^4}
J^r_{KK}(s)+\frac{m_\pi^4}{6 f_\pi^4}J^r_{\eta\eta}(s)
\end{equation}

	It is straightforward to see that the imaginary part of eq.
(\ref{pipi1}) is the one required by unitarity up to $\mathcal{O}(p^4)$ 
for the I=0 S-wave $\pi\pi$ 
elastic partial wave with pions, kaons and etas as intermediate states. The
squared amplitudes in front of the loop functions are the lowest order $\chi$PT
amplitudes since loops are calculated at $\mathcal{O}(p^4)$. This is the same
kind of result we would obtain for the loop contributions in the s-channel 
from the expansion of the generalization of eq. (\ref{TL}) to coupled channels up to the order 
considered in eq. (\ref{pipi1}), after dividing eq. (\ref{pipi1}) by a global 
factor 2  to match with our normalization in eq. (\ref{identical}) with
$\alpha=2$. However, as we discussed above we use $(-)g_0(s)_{ii}$ instead of
$J^r_{ii}(s)$ in eq. (\ref{pipi1}) in order to evaluate the loop contributions in
the s-channel. Hence, we must include in $\hbox{T}_{Left}$ the following
expression 

\begin{equation}
\label{contri}
\frac{(2s-m_\pi^2)^2}{2 f_\pi^4} (J^r_{\pi\pi}(s)+g_0(s)_{11})+
\frac{3 s^2}{8 f_\pi^4}(J^r_{KK}(s)+g_0(s)_{22})
+\frac{m_\pi^4}{6 f_\pi^4}(J^r_{\eta\eta}(s)+g_0(s)_{33})
\end{equation}
 
 The former contribution, together with the exchange of resonances and loops in
the crossed channels, after projecting over the S-wave using eq.
(\ref{identical}) with $\alpha=2$, define $\hbox{T}_{Left}$ in our approach.

	For the I=1/2, $L$=0 $K\pi$ partial wave one has 
essentially the same situation than for $\pi\pi$. From eqs.(3.6) and (3.19) 
one calculates the contribution of loops, which we project over the S-wave. The
loops in the s-channel give the corresponding result to eq. (\ref{pipi1}) for 
the $K\pi$ I=1/2 S-wave. In this case, instead of having the loop
function $J^r_{PP}(s)$, one can write it in terms of $\bar{J}_{K\pi}(s)$ and 
$\bar{J}_{K\eta}(s)$. After taking into account the difference between 
$\bar{J}_{PQ}(s)$ and our loop functions, 
one obtains the analog result to eq. (\ref{contri}) for $K\pi$. This 
contribution, together with the projection over the S-wave of loops and 
the exchange of resonances (eqs. (3.6), (3.20)) in crossed channels, give 
$\hbox{T}_{Left}$. 

We have not considered the tadpole contributions coming from pseudoscalar loops
without flux of energy and the coupling of scalar resonances to the vacuum (
Fig. 2.b of \cite{UMR}) because they are reabsorbed into the residues and
positions of the CDD poles and the subtraction constant $a_0$, eq.(\ref{fin/d}), 
which we have phenomenologically fixed.

\begin{table}
\begin{center}
\caption{Influence of the unpysical cuts for the I, $L$=0 $\pi\pi$ and I=1/2,
$L$=0
$K\pi$ partial waves. The three first columns refer to $\pi\pi$ and the last
three to $K\pi$.}
\vspace{0.2cm}
\begin{tabular}{c|c|c||c|c|c}
\hline
$\sqrt{s}$  & $\frac{\hbox{T}_{Left}}{|\hbox{T}_{0,11}|}$
& $\frac{\hbox{T}_{Left}}{\hbox{T}^\infty_{0,11}}$ & $\sqrt{s}$  & $\frac{\hbox{T}_{Left}}{|\hbox{T}_{0,11}|}$
& $\frac{\hbox{T}_{Left}}{\hbox{T}^\infty_{0,11}}$  \\
MeV & $\%$ & $\%$ & MeV & $\%$ & $\%$\\
\hline
276 & 3.7 & 4.8 &634 & 7.1 & 8.7 \\
376. & 3.5 & 5.1 &684 & 3.7 & 4.7 \\
476 & 4.1 & 5.7 &734 & 0.3 & 0.4 \\
576 & 5.7 & 6. &784 & -2.5 & -3.3 \\
676 & 8.1 & 6.1 &834 & -5.7 & -7.2 \\
776 & 11.2 & 5.6 &   & & \\
\hline
\end{tabular}
\end{center}
\end{table}

	Including explicit resonance fields as done in \cite{UMR} increases the
range of safe applicability of Chiral Symmetry from $\sqrt{s}\approx$400 
MeV, accomplished in $\chi$PT, up to $\sqrt{s}\approx 700-800$ MeV, as can be 
seen in \cite{UMR} when 
comparing their results with the experimental data. 

The results which we obtain for the contribution of $\hbox{T}_{Left}$ in the
range of energies of \cite{UMR} are shown in Table 3. In the second 
and fifth columns we show, respectively, the ratio between 
$\hbox{T}_{Left}$ and the absolute value of our calculated I, $L$=0 $\pi\pi$ 
and
 I=1/2, $L$=0 $K\pi$ partial wave amplitudes up to $\sqrt{s}\approx 800$ MeV. In
 Table 3 we also compare $\hbox{T}_{Left}$ with the tree level amplitudes
$\hbox{T}^\infty_{0,11}$. This ratio is also significative because the procedure
which we have followed to arrive to a unitarized amplitude from 
$\hbox{T}^\infty_{0,11}$ would not be much affected by the addition of 
$\hbox{T}_{Left}$ which is a small correction with respect to 
$\hbox{T}^\infty_{0,11}$. We 
 see that these ratios are rather small. Therefore, this supports our point 
 of view of treating the left hand cut as a perturbation in the range of 
energies we have considered.

	It is worth mentioning that this smallness of the unphysical cuts, as 
shown in Table 3, is a consequence of a cancellation between the contributions 
to T$_{Left}$ from the loops and the exchange of resonances in crossed channels. 
In fact, the individual contributions in the $\pi\pi$ case, for energies around 
$\sqrt{s}=600$ MeV, are of the order of 15-20$\%$ with respect to
$\hbox{T}^\infty_{0,11}$.  	

In a recent work, ref. \cite{Igi}, the authors also combine the N/D method with chiral
symmetry studying the (I,$L$)=(0,0), (2,0) and (1,1) $\pi\pi$ partial wave
amplitudes. However, in this work only elastic unitarity is considered and the
calculations are done in the chiral limit ($m_\pi=0$). On the other hand,
the left hand cut is approximated only by the exchange of the $\rho$ plus a scalar
resonance without including loops in the crossed channels. These loops, as we have
seen in this section, cancel to a large extend the crossed resonance contributions for
the S-wave I=0 $\pi\pi$ scattering. 

\section{Conclusions}

Making use of the $N/D$ method, we have developed the most general structure
 that an elastic partial wave amplitude has when the unphysical cuts are 
 neglected. After matching this result with lowest order, ${\mathcal{O}}(p^2)$, 
 $\chi PT$ 
 \cite{Retaila} and with the exchange of resonances with spin $\leq$1, in a way
consistent with chiral symmetry as given in ref. \cite{EP}, we extend the 
formalism to handle also coupled channels. Then, $\pi\pi$ 
and $K\pi$(I=1/2) P-wave amplitudes are described up to $\sqrt{s}=1.2$ GeV. It
 is shown that these amplitudes can be given rather accurately in terms of 
$m_\pi$, $m_K$, $f$ and the masses of the $\rho$ and $K^*$ resonances, 
when restrictions coming from Large $N_c$ QCD and unitarity are considered, 
in the lines of what was observed in \cite{FP}. 

Next, the scalar sector is studied and good agreement with
experiment up to $\sqrt{s}=1.4$ GeV is found. An octet and a
singlet are included with masses around 1.4 and 1 GeV respectively. The former
 originates the observed $f_0(1500)$, $a_0(1450)$ and $K^*_0(1430)$ resonances,
  the latter
an important contribution to the physical pole of the $f_0(980)$. Other
poles appearing in our amplitudes, the $a_0(980)$, $\sigma$, $\kappa$ and an
important contribution to the final $f_0(980)$, originate from meson-meson
scattering with the lowest order $\chi PT$ amplitudes plus the constant $a^{SL}$
as dynamics source. This situation is very different from the
 one observed in the vector channels where tree level structures dominate
 the scattering process and a strong suppression of unitarity loops occurs, as 
 indicated at the end of {\bf{section 4}}. As a consequence, the present study
 supports that a concept like scalar meson dominance, analogous to the well
 known vector meson one, is not suited at the phenomenological level.
 
 In the last section we have made some estimations in order to investigate the
influence of the unphysical cuts. The results obtained support our
 picture of treating the unphysical cuts in a perturbative way and then
 establishing the stability of our conclusions in {\bf{sections 3}} and 
 {\bf{4}} against the corrections coming from cross symmetry.

 \vspace{1cm}
 {\bf{Acknowledgments}}

 We would like to acknowledge fruitful and basic discussions for the present
 work with A. Pich. Discussions with S. Peris, J.R. Pel\'aez and A. Kataev 
 are also acknowledged. This work was partially supported by DGICYT under 
 contacts PB96-0753 and by the EEC-TMR Program$-$Contact No. ERBFMRX-CT98-0169. 
 J. A. O. acknowledges financial support from the Generalitat Valenciana.  

 \vspace{1cm}
\begin{appendix}
\section{N/D in coupled channels}

In this appendix we make use of a matrix formalism to
deal with several coupled channels. In analogy with the elastic case, eq.
(\ref{T'}), let us define the matrix $\hbox{T}'_L$ as

\begin{equation}
\label{A.1}
\hbox{T}'_L(s)=p^{-L} \hbox{T}_L(s) p^{-L}
\end{equation}
with $p$ a diagonal matrix which elements are $p_{ij}=p_i \delta_{ij}$ where
$p_i$ is the modulus of the CM momentum of the channel $i$,
$\displaystyle{p_i=\frac{\lambda^{1/2}(s,m_{1i}^2,m_{2i}^2)}{2\sqrt{s}}}$, with
$m_{1i}$ and $m_{2i}$ the masses of the two mesons in channel $i$.

From the beginning we neglect the unphysical cuts. As a consequence 
$\hbox{T}_L(s)_{ij}$ will be proportional to $p_i^L p_j^L$. This makes that
$\hbox{T}_L(s)_{ij}$, apart from the right hand cut coming from unitarity (above
the thresholds for channels $i$ and $j$, $s_{th}^i$ and $s_{th}^j$ 
respectively), will have another cut for odd $L$ between $s_{th}^i$ and $s_{th}^j$ due to
the square roots present in $p_i$ and $p_j$. In this way, $\hbox{T}'_L$ will be
free of this cut and will have only the right hand cut coming from unitarity.
Thus it will satisfy

\begin{equation}
\label{A.2}
\hbox{Im T}'^{ -1}_L(s)=-p^L \rho(s) p^L=-\rho(s) p^{2L}
\end{equation} 
where $\rho(s)$ is a diagonal matrix defined by

\begin{equation}
\label{A.3}
\rho(s)=-\frac{p}{8\pi\sqrt{s}}\theta(s)
\end{equation}
with $\theta(s)$ another diagonal matrix such that $\theta(s)_{ii}$=1 above the
threshold of channel $i$ and 0 below it.

We write $\hbox{T}'_L$ as a quotient of two matrices, $\hbox{N}_L$ and 
$\hbox{D}_L$ making use
of the coupled channel version of the N/D method \cite{Bjorken}

\begin{equation}
\label{A.4}
\hbox{T}'_L=\hbox{D}'^{ -1}_L \hbox{N}'_L
\end{equation}

We can always take $\hbox{N}'_L$ free of poles and also containing all the zeros
of $\hbox{T}'_L$. In such a case $\hbox{N}'_L$ will be just a matrix of
polynomials, we then write

\begin{equation}
\label{A.5}
\hbox{N}'_L=\hbox{Q}_{n-L-1}
\end{equation}
with $\hbox{Q}_{n-L-1}$ a matrix of polynomials of maximum degree $n-L-1$.

In this way, from eq. (\ref{A.2}) and (\ref{A.4}) one has

\begin{equation}
\label{A.6}
\hbox{Im D}'_L(s)=-\hbox{N}'_L(s) \rho(s) p^{2L}
\end{equation}
and making a dispersion relation for $\hbox{D}'_L$ one has 

\begin{equation}
\label{A.7}
\hbox{D}'_L(s)=-\frac{(s-s_0)^n}{\pi}\int_0^\infty ds'
\frac{\hbox{Q}_{n-L-1}(s') \rho(s') p^{2L}(s')}{(s'-s)(s'-s_0)^n}+\hbox{P}_{n-1}
\end{equation}
with $\hbox{P}_{n-1}$ a matrix of polynomials of maximum degree $n-1$.

Because $\hbox{N}'_L$ is just a matrix of polynomials, it can be reabsorbed in
$\hbox{D}'_L$ to give rise to a new $\widetilde{\hbox{D}}'_L$ which will fulfill
eq. (\ref{A.6}) but with $\widetilde{\hbox{N}}'_L=1$. In this way 

\begin{eqnarray}
\label{A.8}
\hbox{T}'_L&=&\widetilde{\hbox{D}}'^{-1}_L \nonumber\\
\widetilde{\hbox{N}}'_L&=&1 \nonumber\\
\widetilde{\hbox{D}}_L'&=&-\frac{(s-s_0)^L+1}{\pi}\int_0^\infty
\frac{\rho(s')p^{2L}(s')}{(s'-s)(s'-s_0)^{L+1}}+\hbox{R}(s)
\end{eqnarray}
with $\hbox{R}(s)$ a matrix of rational functions whose poles will contain the
zeros of $\hbox{T}'_L$. This fact is in clear analogy with the role played by 
the CDD poles included in {\bf{Section 2}} for the elastic case.

\end{appendix}

\end{document}